\documentclass[a4paper,fleqn,usenatbib]{mnras}

\usepackage[T1]{fontenc}
\usepackage{ae,aecompl}

\usepackage{graphicx}	
\usepackage{amsmath}	
\usepackage{amssymb}	
\usepackage{natbib}
\usepackage{float}
\usepackage[usenames,dvipsnames]{xcolor}
\usepackage[]{hyperref}

\usepackage[normalem]{ulem}

\usepackage{times}
\usepackage{url}
\usepackage{subfig}

\newcommand{\vmax}{V_{\rm max}}
\newcommand{\kms}{\mathrm{km~s^{-1}}}
\newcommand{\cmg}{\mathrm{cm^2~g^{-1}}}

\newcommand{\msun}{M_\odot}

\newcommand{\mvir}{M_\mathrm{v}}

\newcommand{\sigmam}{\sigma/m}

\title{A Testable Conspiracy: Simulating Baryonic Effects on Self-Interacting Dark Matter Halos}

\author[Elbert et al.]{Oliver D. Elbert$^1$\thanks{$\!$oelbert@uci.edu},
	James S. Bullock$^1$,  Manoj Kaplinghat$^1$, Shea Garrison-Kimmel$^2$, \and 
	Andrew S. Graus$^1$,  Miguel Rocha$^3$\\
	\noindent$\!$ $^1$Center for Cosmology, Department of Physics and Astronomy,
	  University of California, Irvine, CA 92697, USA \\
	\noindent$\!$ $^2$TAPIR, California Institute of Technology, Pasadena, CA 91125, USA \\
	  University of California, Irvine, CA 92697, USA \\
	  \noindent$\!\!$ $^3$SciTech Analytics, Inc., Santa Cruz, CA, 95062, USA\\}

\begin{document}

\pagerange{\pageref{firstpage}--\pageref{lastpage}} 
\pubyear{2016}

\maketitle
\date{\today}
\label{firstpage}

\begin{abstract}
We investigate the response of self-interacting dark matter (SIDM) halos to the growth of galaxy potentials using idealized simulations, each run in tandem with standard collisionless Cold Dark Matter (CDM). 
We find a greater diversity in the SIDM halo profiles compared to the CDM halo profiles.
If the stellar gravitational potential strongly dominates in the central parts of a galaxy, then SIDM halos can be as dense as CDM halos on observable scales. For extreme cases with highly compact disks core collapse can occur, leading to SIDM halos that are denser and cuspier than their CDM counterparts. If the stellar potential is not dominant, then SIDM halos retain constant density cores with densities far below CDM predictions. 
When a disk potential is present, the inner SIDM halo becomes {\em more flattened} in the disk plane than the CDM halo
These results are in excellent quantitative agreement with the predictions of Kaplinghat et al. (2014).
We also simulated a galaxy cluster halo with a central stellar distribution similar to the brightest central galaxy of the cluster A2667. A SIDM halo simulated with cross section over mass $\sigma/m = 0.1\ \cmg$ provides a good match to the measured dark matter density profile of A2667, while an adiabatically-contracted CDM halo is denser and cuspier. 
The cored profile of the same halo simulated with $\sigma/m = 0.5\ \cmg$ is not dense enough to match A2667. Our findings are in agreement with previous results that $\sigma/m \gtrsim 0.1 \, \cmg$ is disfavored for dark matter collision velocities in excess of about 1500 km/s.  More generally, the predictive cross-talk between baryonic potentials and SIDM density distributions offers new directions for constraining SIDM cross sections in massive galaxies where baryons are dynamically important.
\end{abstract}


\section{Introduction}
\label{sec:intro}

The dark matter (DM) paradigm has been tremendously successful  in explaining the large-scale structure of our universe \citep[see, e.g. ][]{PlanckCos15, SDSS16},  though the precise nature of dark matter remains unknown.  The simplest example of cold dark matter (CDM), consisting of a single, collisionless particle with negligible primordial thermal velocity dispersion,  can match the large-scale data remarkably well.  Alternatively, dark matter could be more complex, with nongravitational coupling to standard model particles
\citep[e.g.][]{BoehmSchaeffer05,Escudero15} and/or new dark sector particles \citep[e.g.][]{Feng10,Khlopov10,Lesgourgues16}; many models of this kind produce observable signatures in astronomical data sets \citep{Mangano06,Feng09,Racine15}.  In this paper we consider the possibility that dark matter has strong elastic self-scattering interactions and explore the implications of such interactions on the dark matter distributions within individual galaxies.  We specifically focus on the back-reaction associated with galaxy formation.

Collisional or Self-Interacting dark matter (SIDM) was first explored in the context of galaxy formation by \citet{Spergel00}, who argued that SIDM models with cross-section over mass $\sigmam \sim 1\ \cmg$ should lead to observable constant density cores in galaxies, in agreement with observations at that time.  While early estimates suggested that SIDM models of this kind would significantly reduce substructure counts compared to CDM, more recent numerical investigations have shown that the substructure differences are minimal \citep{Vogelsberger12,Rocha13}.  However, the original expectation that SIDM halos should have constant-density cores has been demonstrated robustly in cosmological simulations  \citep[][]{Dave2001,Rocha13,Zavala13}.

SIDM cores are generated by energy-exchange interactions, which heat the halo center until it becomes isothermal.  The radial extent of this core is set by the requirement that a typical dark matter particle will experience at least one interaction per Hubble time \citep{Rocha13}.  This implies that larger SIDM cross sections  produce larger isothermal cores.  If the cross section is large enough, the isothermal region can extend beyond the peak in the halo's velocity dispersion profile; in this case, energy-exchange interactions could extract heat from the core leading to core-collapse, which increases the central density \citep{Kochanek2000,Balberg02,Colin02,Koda11,Vogelsberger12}.  However, this effect is muted in cosmological simulations and \citet{Elbert15} used dark matter (only) zoom cosmological simulations to show that core-collapse behavior sets in only for very large cross sections $\sigmam \gtrsim 10\ \cmg$.  

\begin{table*}
\centering
\begin{tabular}{lccccccccccc}
Name & $\mathrm{\mvir}$ & $c_{\rm v}$ & $N_\mathrm{p}$ & $\mathrm{r_{cut}}$ & $\epsilon$ & Convergence Radius & Potential Shape & $\mathrm{M_{\rm gal}}$ &  $\mathrm{\textit{a}}$ & $\mathrm{\textit{b}}$   & $h$ \\ 
		 &	  ($10^{12} \msun$) &  & ($10^6$) & (kpc) & (kpc) & (kpc) & & ($10^{10} \msun$) & (kpc) & (kpc) & (kpc) \\ \hline \hline
MW & $1.0$  & 13 &  $3$ & 230 & $0.4$ & 0.83 & MN Disk & $5.0$ & $1.5, 3.0, 6.0$ & $0.3$ & -- \\ \hline
LSB & $0.2$  & 11.8 & $10$ & 170 & $0.19$ & 0.30  & MN Disk & $0.06$ & $2.2$ & $0.2$ & --  \\ \hline
Elliptical & $1.8$ & 9.7  & $10$ & 300 & $0.37$ & 1.0  & Hernquist Sphere & $6.2$ & $ - $ & -- & 3.0 \\ \hline
Cluster & $10^3$ & 3  & $50$ & 500 & $2$ & 3.4  & Hernquist Sphere & $210$ & $ - $ & -- & 28.5 \\ \hline
\end{tabular}
\caption{Summary of simulated halos. The first five  columns list identifying 
names and general simulation properties: halo mass, NFW concentration, particle number, exponential cutoff radius and force softening.   We define $\mvir$ following \citet{Bryan98} with a virial overdensity of $\Delta_{\rm v} = 97$ with respect to the critical density.  The sixth column lists the convergence radius for the SIDM runs, which we set to $0.6$ times the \citet{Power03} radius for CDM as found in Elbert et al. (2015).
The last four columns summarize the properties of the galaxy potentials grown in each case, where $M_{\rm gal}$ is the final galaxy mass and the other shape parameters are defined in Equations \ref{eqn:MN} and \ref{eqn:H}.  Note that there are three separate disks of varying scale length for the Milky Way runs.  We refer to these in the text and figures as  `Compact', `Fiducial', and `Extended.'}
\label{sims.tab}
\end{table*}

The tendency for SIDM models with $\sigmam \lesssim 10\ \cmg$ to produce constant-density cores with lower overall density 
is of special interest for comparisons to dwarf and low surface brightness (LSB) galaxies.  This is because many of these galaxies are observed to have 
cores on roughly the scales expected in SIDM \citep{Flores1994,Moore94,deBlok96,Salucci00,deBlok01,Swaters03,Gentile04,Simon05,Spekkens05,KuziodeNaray08,deBlok08,Donato09,Oh11,Adams14} 
as opposed to the cusps predicted in dissipationless CDM simulations \citep{Dubinski1991,NFW}.  SIDM cores 
also may provide a natural explanation for the unexpectedly low densities of local dwarf galaxies \citep{Vogelsberger12,Vogelsberger14,Elbert15}, 
a problem known as  ``Too Big to Fail" (TBTF)  \citep{MBK11,MBK12,Ferrero2012,Klypin2014,Papastergis14,GarrisonKimmelTBTF,Tollerud14}.  
There are many in the galaxy formation community who believe these issues may be resolved by baryonic processes such as supernova feedback 
\citep{Navarro96,Read05,Pontzen12,Governato12,DiCintio14,Onorbe15,Maxwell15,Dutton16,Katz16,Read2016} though not all authors necessarily agree \citep{Penarrubia12,SGK13,Pace16}.  Tidal effects have been shown to solve TBTF in satellite galaxies \citep[see e.g.][]{Read06,Zolotov2012,Brooks2014,DelPopolo2014,Arraki14}, but the evidence for TBTF in the local field \citep{Kirby14,GarrisonKimmelTBTF} necessitates another solution for these galaxies.
This ongoing debate and the lack of DM detections in direct, indirect and collider searches motivates a thorough exploration of the SIDM hypothesis.

The goal of this paper is to investigate the effects of galaxy formation on SIDM halos, specifically the contraction of these halos due to the gravitational potential of the galaxy.  To this end we use a set of N-body simulations similar to those initially used to examine contraction in CDM halos.  The work is organized as follows: in \S \ref{sec:constraints}, we briefly describe the properties required of a viable SIDM model and in \S\ref{sec:motivation}, we sketch the physics of contraction of SIDM halos and motivate our work in this paper. We describe our simulations and analysis in \S\ref{sec:simulations}.  We present our results in \S\ref{sec:results}, discussing our Milky Way analogue halos in \S\ref{ssec:MW} and our elliptical and lsb simulations in \S\ref{ssec:other}, while in \S\ref{ssec:analytic} we directly compare our simulations to the analytic model presented in \citet{Kaplinghat15b}.  \S\ref{sec:cluster} shows the results of our cluster simulations, and compares these with the observations of \citet{Newman13b}.  We summarize our results and conclude in \S\ref{sec:conclusions}.

\section{Properties of viable SIDM models}
\label{sec:constraints}

Previous work has placed constraints on the SIDM cross section over mass across a range of halo masses.  Generally, $\sigmam$ below $0.1\ \cmg$ have been found to be indistinguishable from CDM models \citep{Rocha13}.  In low mass galaxies with maximum circular velocity $\vmax \simeq 30\ \kms$, $\sigmam$ values ranging from 0.5 to 10 $\cmg$ alleviate the core-cusp and TBTF \citep{Elbert15,Zavala13,Vogelsberger12,Fry15}.  However, values significantly in excess of 1 $\cmg$ may lead to efficient tidal stripping of stars in the satellites of the MW and Andromeda \citep{Gnedin01,Penarrubia10,Dooley16} , providing a possible avenue for an upper limit on the cross section in the future. 
Recent work by \cite{Kaplinghat15b} showed that SIDM models with cross sections around 2 $\cmg$ fit the rotation curves of the 12 analyzed dwarf and low surface brightness galaxies well. This work used an analytic model built on arguments discussed previously \citep{Rocha13,Kaplinghat14b} and showed that the analytic model is a good match for the density profiles of halos in DM-only cosmological SIDM simulations. The summary of above constraints is that for collisional velocities of order 100 km/s or smaller, a $\sigmam$ values close to 1 $\cmg$ is favored and consistent with all existing constraints. A larger sample of rotation curves will reduce uncertainties in the determination of the cross section on galactic velocity scales. 

The dark matter velocities in galaxy clusters are an order of magnitude larger than in dwarf galaxies, and many techniques have been used to constrain the DM self-interaction cross section at these velocities.  Cluster mergers have been used by many studies to constrain SIDM \citep[see e.g.][]{Randall08,Dawson12,Kahlhoefer14,Massey15,Schaller15,Kim16,Robertson16}, with typical limits of $\sigmam \leq 1\ \cmg$ on the self-interaction cross section. However, recent work by \citet{Kim16} showed that constraints based on the displacement of the stellar and dark matter centroids are overly stringent, weakening previous constraints. They find that the displacement of the brightest cluster galaxy (BCG) relative to the halo center may be a better observable, possibly allowing $\sigmam$ values around $0.1\ \cmg$ to be tested.  The Bullet cluster constraint based on the mass loss from the merging sub-cluster \citep{Markevitch2004} also needs to be reevaluated using self-consistent SIDM merger simulations and taking into account cosmic variance in the initial conditions. In addition, we also need theoretical refinement to apply these constraints to velocity-dependent cross sections. 
Cluster shapes have provided an orthogonal method of investigating self-interaction cross section on these scales. Core formation in SIDM halos leads to more spherical inner density profiles, so measurements of the ellipticities of cluster halos have been used to constrain $\sigmam$ values to below $0.1\ \cmg$ \citep[][]{Miralda-Escude02}. However, due to the large scatter in axis ratios, the ability of SIDM halos to retain some triaxiality, and the observational methods used to constrain halo shapes, the ellipticity constraints are unlikely to be better than about $1 \ \cmg$ for cluster velocities \citep[see][for a detailed discussion]{Peter13}. 

The most stringent constraint on cluster velocity scale arises from the fact that the measured dark matter density profiles are in substantial agreement with CDM outside the half-light radii of the BCGs. Within about 10-50 kpc (range of BCG half-light radii), however, the dark matter density profile is shallower than the CDM expectations \citep{Newman13b}, as we discuss later. \citet{Kaplinghat15b} used these measurements to show that the preferred cross section for relative velocities larger than 1000 km/s is about 0.1 $\cmg$, consistent with earlier results from \citet{Yoshida00b}. If the inferred shallowness of the density profile is due to AGN feedback or some other baryonic process \citep[e.g.]{Martizzi2013}, then this value of 0.1 $\cmg$ provides a stringent upper limit on the self-interaction cross section for velocities in excess of 1000 km/s. This result and the large difference in DM velocities in dwarf galaxies and galaxy clusters demand that viable SIDM models must have a velocity dependent self-interaction cross section. The required velocity dependence -- from 1 $\cmg$ for velocities below about 100 km/s to 0.1 $\cmg$ for velocities above 1000 km/s  -- can be easily accommodated in a variety of particle physics models \citep{Feng10b,Loeb11,Tulin13,Cline14,Boddy14b,Kwa16}. 

\section{Motivation: contraction of SIDM halos}
\label{sec:motivation}

Our work on simulating SIDM halos including a baryonic component is important for two specific reasons. First, in galaxies dominated by baryons there is no systematic evidence for large cores or lowered dark matter density profiles \citep[e.g.][]{Cappellari15}. Second, in galaxy clusters the stars dominate the total mass budget within their half-light radii, yet the dark matter tends to be under-dense compared to predictions. There is a simple analytic model (which we discuss later) that can explain both these observations but it has not been tested against simulations including a stellar component. By testing the accuracy of the analytic model we are able to bolster the case for a velocity dependent self-interaction cross section. We also test the possibility of core collapse in systems that have extremely dense baryonic distributions. 

Since our aim is to test and further elucidate the physics of how halos become isothermal in the potential well of the baryons, we have chosen to run idealized simulations with disks grown adiabatically. This is complementary to the approach of running full-fledged hydrodynamic simulations that include self-interactions between dark matter particles. \citet{Vogelsberger14} and \citet{Fry15} examine dwarf galaxies ($\mvir \sim 10^{10}~\msun$) using fully self-consistent hydrodynamic simulations and find that observable cores are still formed in these dwarfs in SIDM.  \citet{Fry15} find that the cores in their simulated dwarfs are not substantially different from those formed purely via feedback in their CDM simulations.  It is not clear how these results generalize to larger halo masses, where feedback is expected to be less important in driving core formation in CDM halos and halo contraction effects are expected to dominate \citep[e.g.][]{DiCintio14,Dutton16,Fiacconi16}.  

The fact that growing baryonic potentials can cause contraction of CDM halos was first investigated analytically by \citet{Blumenthal86} and \citet{Ryden&Gunn87}, who used an adiabatic invariant approach.  They demonstrated that if a baryonic potential grew to dominate the central potential, the entire dark halo would contract, increasing the central dark matter density by more than an order magnitude in plausible cases.  Other studies \citep[e.g.][]{Jesseit02} ran numerical simulations of isolated DM halos with disk potentials and found the DM density in these halos reproduced analytical predictions.  \citet{Gnedin04} studied baryonic contraction in hydrodynamic cosmological simulations and found that both adiabatic model predictions and isolated simulations produced central densities that were roughly $50\%$ too high in halos where baryons dominate and proposed an alternative model to encapsulate the adiabatic contraction effect.  

Though previous work has examined adiabatic contraction in collisionless DM halos, they show that it typically occurs at early times when the disk or bulge of a galaxy is forming. Moreover, the adiabatic contraction timescales are much shorter than the typical timescale for self-interactions in galaxies, assuming a cross section around 1 $\cmg$. Thus, we expect adiabatic contraction to happen (at early times) in SIDM halos as in CDM halos, unless inhibited by feedback. The baryon dominated systems we simulate are Milky Way size or larger and we assume that feedback does not prevent adiabatic contraction in these systems \citep{DiCintio14,Dutton16}. Over longer timescales, the self-interaction process will allow the halo to become isothermal out to a certain radius and this process can make the dark matter density profile shallower or retain the steep density profiles created by adiabatic contraction or steepen it further in the case of core-collapse. The outcome depends on a comparison of the gravitational potential of the baryons to the velocity dispersion of dark matter. 

\citet{Kaplinghat14b} discussed the response of SIDM halos to the formation of a stellar disk or bulge using analytic equilibrium models.  They found that the resultant SIDM core radius and density should be linked closely to the underlying baryonic potential in systems where the baryons are important dynamically. In the limit where the stars dominate the gravitational potential, the dark matter density profile (in the region where it is isothermal) scales as $\exp(-\Phi_\star(r)/\sigma_{v0}^2)$ \citep{Amorisco10,Kaplinghat14b} where $\Phi_\star$ is the gravitational potential of the stars and $\sigma_{v0}$ is the 1-d central velocity dispersion of dark matter. Thus stars and dark matter are tied together in terms of the density profile and the shape of the dark matter halo must follow the contours of the stellar gravitational potential. This result emphasizes the need to account for baryonic processes when exploring SIDM phenomenology in galaxies with significant gas or stellar components.  

The goal of this paper is to numerically investigate the effects of baryonic contraction on SIDM halos using a set of isolated N-body dark matter simulations of Milky Way, Elliptical, LSB, and Cluster analogue halos.  These simulations are similar in spirit to those first used to test and confirm analytic contraction models in the context of CDM.    
All of our halos were simulated with fixed SIDM cross sections of  $\sigma/m = 0.5\ \cmg$, but we show below that our simulations reproduce the model of \citet{Kaplinghat15b}, indicating we can use their model to extend our results to a wide range of cross sections. 

\section{Simulations}
\label{sec:simulations}

Our code is a modified version of \texttt{GADGET-2} \citep{Springel05} that allows for the inclusion of hard-sphere scattering between dark matter particles \citep{Rocha13}.   The simulations consist of a series of $3$ to $50$ million particle dark matter halos, initialized as \citet*{NFW} profiles, and run in isolation with and without  an analytic galaxy potential.  The potentials are grown linearly in time from a mass of zero to a final mass $M_{\rm gal}$ in 1 Gyr at the start of our simulations.  We also simulated our Milky Way halo forming the fiducial disk after the SIDM core stabilized and found no difference in the DM distribution.  The particle initial conditions were generated by the public code \texttt{SPHERIC} \footnote{ \url{https://bitbucket.org/migroch/spheric}}, which was first introduced in \citet{SGK13}.   To increase our effective resolution we exponentially truncate the outer regions of our initial NFW halos.  These truncation radii lie far outside the halo scale radius except in the case of the Cluster, which we truncate aggressively in order to resolve the central few kpc of the halo.

Table \ref{sims.tab} summarizes our simulations, which consist of four characteristic halo/galaxy mass~\footnote{Our virial mass definition follows \citet{Bryan98} for a flat LCDM cosmology with $\Omega_{\rm m} = 0.27$.} combinations:
\begin{enumerate}
\item Milky Way: $\mvir = 10^{12} \msun$ with disks $M_{\rm gal} = 5 \times 10^{10} \msun$;
\item  LSB: $\mvir = 2\times10^{11} \msun$ with disk $M_{\rm gal} = 6 \times 10^8 \msun$; 
\item  Elliptical: $\mvir = 1.8 \times 10^{12} \msun$ with $M_{\rm gal} = 6.2 \times 10^{10} \msun$; and
\item  Cluster: $\mvir = 1 \times 10^{15} \msun$ with $M_{\rm gal} = 2.1 \times 10^{12} \msun$. 
\end{enumerate}
Each of these simulations are run with CDM and SIDM and also with and without the galaxy potentials for comparison.  
The Milky Way halo mass simulations include three separate galaxy disk potential runs, each of which has a fixed galaxy mass but variable scale length (see below).  We present 22 simulations in all.  
All SIDM halos were run with $\sigmam = 0.5\ \cmg$ and used a self-interaction smoothing factor of 25 \% of the force softening length \citep{Rocha13}. We also resimulated our cluster halo with $\sigmam = 0.1\ \cmg$. 
The force softening and convergence radius \citep{Power03,Elbert15} of each run are indicated in columns five and six.
     
For the disk potentials in the LSB and Milky Way runs we adopt the form of \citet{Miyamoto-Nagai75}: 
\begin{equation}
\label{eqn:MN}
\Phi_{\rm MN}(R,z)=\frac{-G M_{\rm gal}}{\sqrt{R^2+(a + \sqrt{b^2+z^2})^2}},
\end{equation}
where $a$ defines a scale length and $b$ sets a scale height.

\begin{figure*}
\centering
\subfloat{\includegraphics[width = .715\columnwidth,trim={1.2cm 0 1cm 0}]{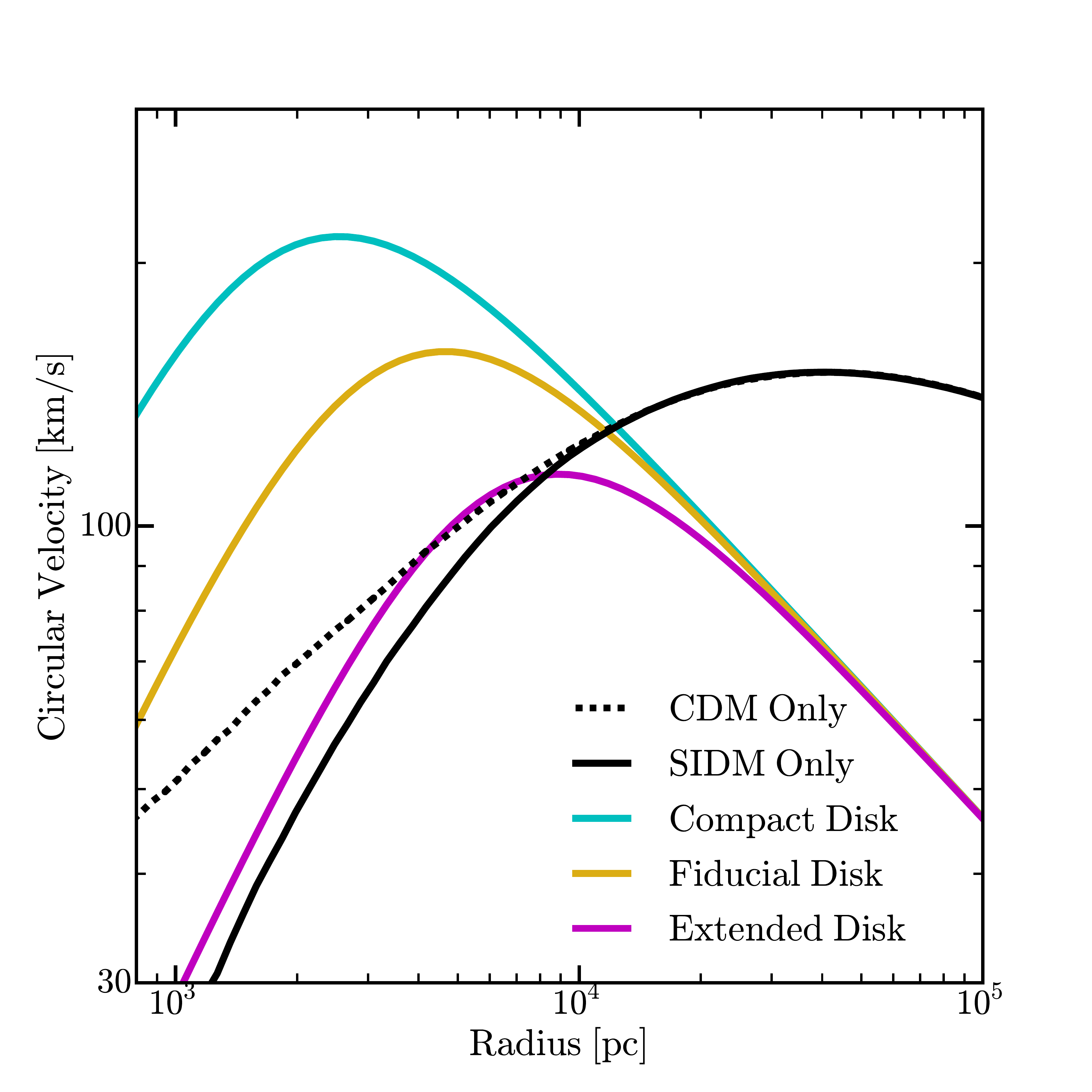}}
\subfloat{\includegraphics[width = .715\columnwidth,trim={1.2cm 0 1cm 0}]{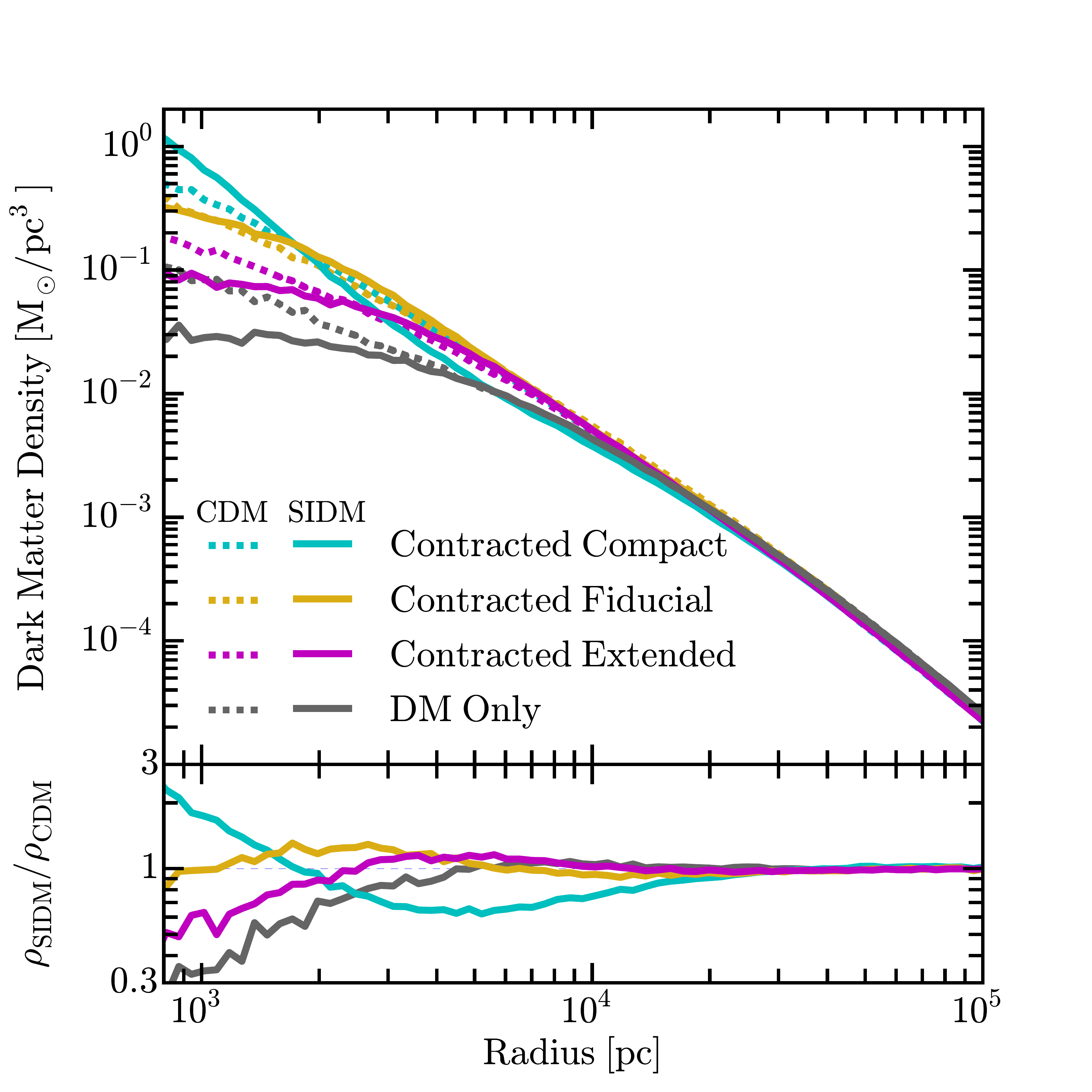}}
\subfloat{\includegraphics[width = .715\columnwidth,trim={1.2cm 0 1cm 0}]{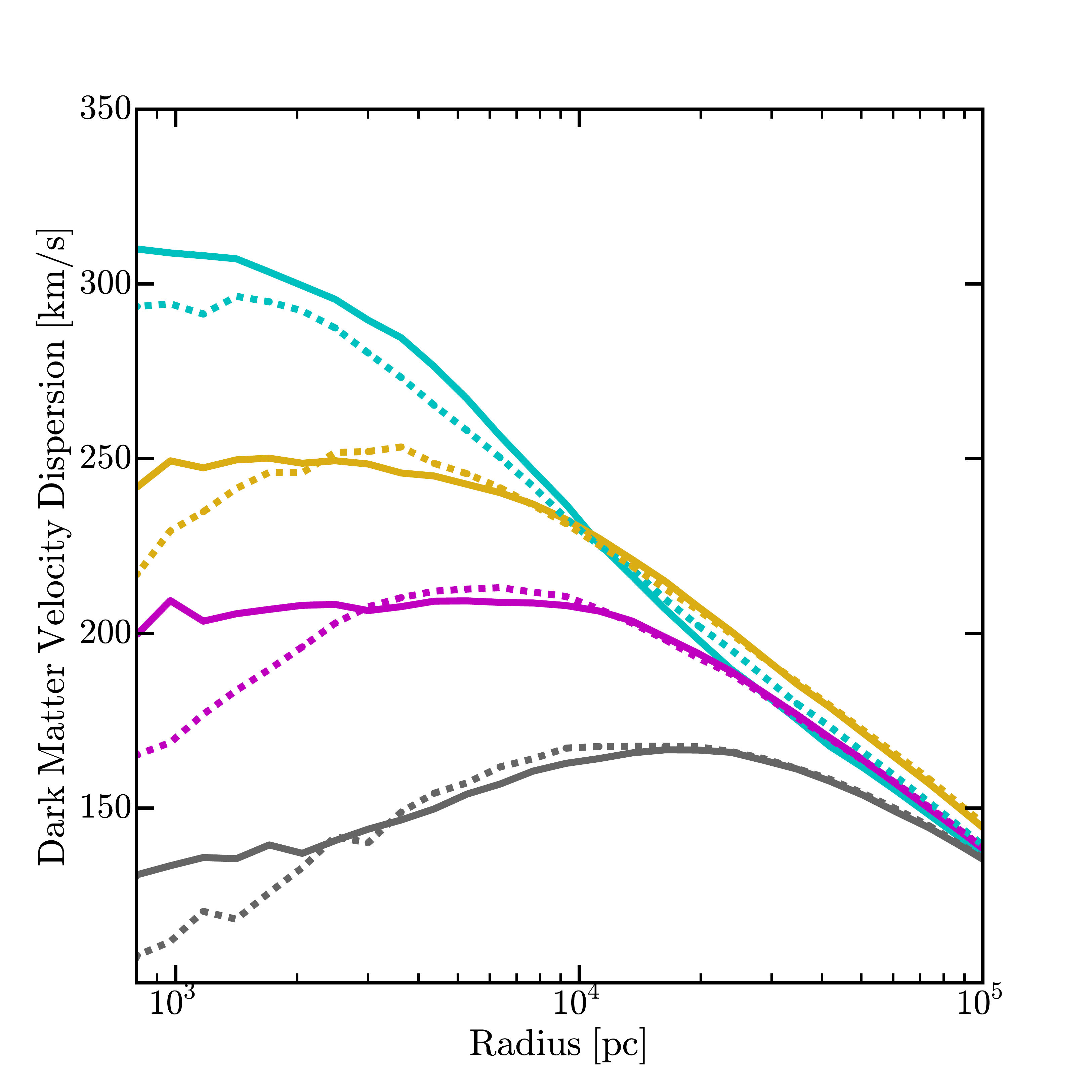}}
\caption{\textit{Left:} The black lines show circular velocity profiles for dark-matter only Milky Way size halos (SIDM, solid; CDM, dashed).  The colored lines show the circular velocity profiles of three imposed baryonic potentials: Fiducial Disk (yellow), Extended Disk (magenta), and Compact Disk (cyan).   \textit{Middle:} Dark matter density profiles without (grey) and with (colored) response to the grown disk potentials.  The lower panel shows the ratio of the SIDM run to CDM run as a function of radius for each set of simulations.  \textit{Right:} Velocity dispersion profiles of the same halos, which demonstrate isothermal cores for the SIDM runs, as expected.   Note that the relative effect of baryonic contraction is much more substantial in SIDM: the central densities at $600$ pc increase by a factor of $\sim 70$ from the non-contracted case to the Compact Disk case in SIDM, compared to only a factor of $\sim 5$ in the CDM case.  Interestingly, the Fiducial Disk runs in SIDM and CDM have very similar normalizations, though the SIDM simulation does show a small core developing within $\sim 800$ pc.  The Compact Disk, on the other hand, has led to core-collapse in the SIDM halo, resulting in a much higher central density than even the contracted CDM halo.  Core collapse is expected when the velocity dispersion has a negative gradient within the scattering radius, as is clearly the case for the Compact Disk in the right panel.}
\label{fig:densprofiles}
\end{figure*}

Each Milky Way simulation has a fixed scale height $b=0.3$ kpc.  We explore
three scale lengths: $a =1.5$, $3.0$, and $6.0$ kpc, which we refer to as `Compact Disk', `Fiducial Disk', and `Extended Disk', respectively. These values~\footnote{
We map the MN disk parameter to quoted exponential disk scale lengths $R_d$ by requiring that the half-mass radii are equal.  This implies $a \simeq 1.25 R_d$ for the range of parameters we explore.} roughly span the lower two-sigma to upper one-sigma of disk sizes for $M_{\rm gal} = 5\times10^{10}\ \msun$ galaxies \citep{Reyes11}.  In particular, the compact disk is extremely dense and was chosen to investigate whether core collapse occurs. Our LSB disk mimics a typical  LSB from \citet{KuziodeNaray08}, with $M_{\rm gal} = 6.3\times10^{8}\ \msun$, $a = 2.2$ kpc, and $b = 0.2$ kpc. 

We use spherical \citet{Hernquist90} distributions for the Elliptical and Cluster galaxy runs:
\begin{equation}
\label{eqn:H}
\Phi_{\rm H}(r)=\frac{-G M_{\rm gal}}{(r + h)}.
\end{equation}
For the Elliptical, we adopt $h= 3.0$ kpc and $M_{\rm gal} = 6.2 \times 10^{10} \msun$ motivated by matching themedian of bin 28 in  \citet{GravesETG2}.  We relate the typical effective radius for galaxies of this size by demanding that the 3D half-light radii are equal:  $h = R_{\rm e} / 1+\sqrt{2}$.  For the central cluster galaxy we match the results for A2667 as quoted in \citet{Newman13a} by fitting a Hernquist profile with the same half-light radius to the best-fit dPIE profile, which yields $M_{\rm gal} = 2.1\times10^{12} \msun$ and $h = 28.5$ kpc.

Each simulation was analyzed after reaching dynamical equilibrium, such that the density profile was no longer evolving.  This occurred within $\sim 5$ Gyr for all cases except the ``compact" SIDM Milky Way, which underwent core collapse and showed a slowly increasing core density until we stopped the simulation after 10 Gyr.

\section{Results}
\label{sec:results}

\subsection{Milky Way Halos}
\label{ssec:MW}

\begin{figure*}
\centering
\includegraphics[width = 2.1\columnwidth,trim={0 10cm 0 0},clip]{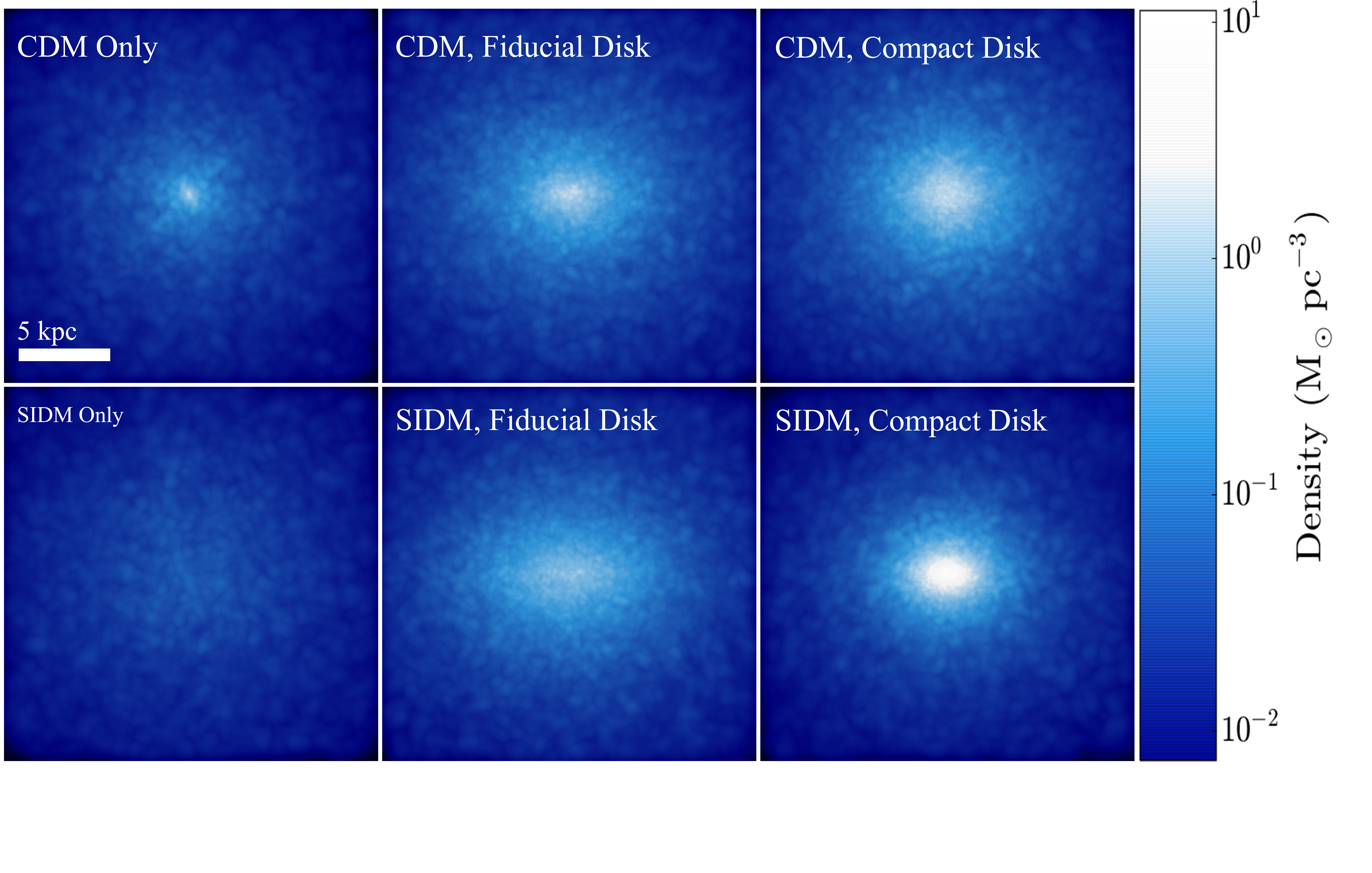}
\caption{Edge-on vizualizations of the dark matter density of an idealized Milky Way sized halo formed with CDM (top) and SIDM (bottom).  Images along the left edge are dark-matter only runs.  The middle panels show the dark matter after the growth of a disk like the Milky Way.  The far right panel shows the dark matter after the growth of a compact disk (see Table 1).   As expected, the SIDM-only simulation has a much lower central density than the CDM-only case.  When a Milky Way-like disk is imposed, both CDM and SIDM halos contract to similar overall central densities, but the SIDM halo tracks the shape of the disk potential more closely than CDM owing to its isothermal velocity distribution.  The compact disk case drives core-collapse in the SIDM simulation, and thus results in an even higher core density than in the contracted CDM run.}
\label{fig:vizes}
\end{figure*}

The setup for the Milky Way analogue simulations is shown in the left column of Figure \ref{fig:densprofiles}.  The black dotted line shows the circular velocity curve ($V_{\rm c}(r) = \sqrt{G M(<r)/r}$) for our CDM-only run (equivalent to the initial conditions) and the solid black line shows $V_{\rm c}(r)$ for SIDM-only run, which is noticeably less dense in the center owing to SIDM core formation \citep[e.g.,][]{Rocha13}.  The colored lines show the implied in-plane circular velocities ($\mathrm{\sqrt{r\frac{d\Phi }{d r}}\vert_{z=0}}$) for the imposed Extended, Fiducial, and Compact Disk potentials.  These show that the disk potential is dominant in our compact and fiducial runs.  Our goal is to explore the halo back-reaction to the growth of each of these components.

The middle column of Figure \ref{fig:densprofiles} shows the dark matter density profiles for all simulations.  The solid lines correspond to the SIDM model and dotted lines to the CDM model.   The gray curves are the dark-matter only runs while the colored lines show what happens after the potentials are grown.    For reference, the bottom panel plots the ratio of dark matter density in SIDM to CDM as a function of radius for each set of runs.

As expected, both CDM and SIDM halos contract in response to galaxy potential growth~\footnote{We have confirmed that our CDM runs generally adhere to the expectations of standard \citet{Blumenthal86} contraction}.   The relative differences are enlightening.    While the dark-matter-only simulations are quite distinct between the cuspy CDM run and the cored SIDM within $\sim 3$ kpc, the dark matter profiles in the Fiducial Disk runs are almost identical down to the resolving limit.  Specifically, the SIDM halo has responded more to the imposed potential than the CDM halo and this has driven the two profiles to a very similar end state.  We only begin to see the formation of an SIDM core within $\sim 1$ kpc, which is similar to the Milky Way core size measured in \citet{Portail16}.  The Extended Disk runs, which impose a less severe potential, have maintained something closer to the original differences, with SIDM beginning to roll off towards a core within $\sim 3$ kpc, but the differences between the SIDM and CDM are less severe than in the DM-only case (which disagree at ~5 kpc).  Finally, the Compact Disk has produced a dramatic change: the SIDM halo is now {\em more dense} than CDM at small radii, with a very cuspy distribution $\rho \sim r^{-2.5}$ at $r \sim 2$ kpc. This is a result of core-collapse: the compact disk potential has heated the dark matter to such an extent that it is now hotter in the core than in the outer part.  The SIDM particles are conducting heat outwards, resulting in a loss of core pressure and subsequent mass inflow.

The SIDM phenomenology is clarified in the right panel of Figure \ref{fig:densprofiles}, which shows the velocity dispersion profile of the dark matter in each run.  The two SIDM simulations with clear constant-density core behavior (DM-only and Extended Disk) are seen to have well-established isothermal velocity distributions at small radii.  In these cases, the SIDM halos are hotter in their cores and colder in their outer regions than their CDM counterparts.  This is exactly the situation that leads to heat transfer from the outside in.   The same effect is seen, though much more mildly, in the Fiducial Disk case.   In the Compact Disk runs, even the CDM halo is hotter in the core than in the outer part.  Such a declining velocity dispersion profile is subject to outward heat flow in the SIDM simulation, and this drives core collapse.

We show in Section~\ref{ssec:analytic} that the resultant density profiles are well explained by the analytical predictions presented in \citet{Kaplinghat14b,Kaplinghat15b}, with the exception of the Compact Disk.  However this is to be expected, given the gravothermal core collapse occurring in the Compact Disk, which violates the assumption of isothermality the model is based on.

Figure \ref{fig:vizes} displays visualizations of the dark matter in three pairs of our simulations, with CDM runs shown along the top and SIDM along the bottom.  The left row shows the dark-matter only versions of each simulation.  The middle and right rows show resultant dark matter distributions after the growth of our Fiducial and Compact disks, oriented such that the disks are seen edge-on.  These results emphasize that the shapes of the SIDM  halos have been altered substantially by the formation of the disk, mirroring the baryonic potential much more closely than the CDM cases within 10 kpc.

\subsection{LSB and Elliptical Halos}
\label{ssec:other}

\begin{figure*}
\includegraphics[width = \columnwidth]{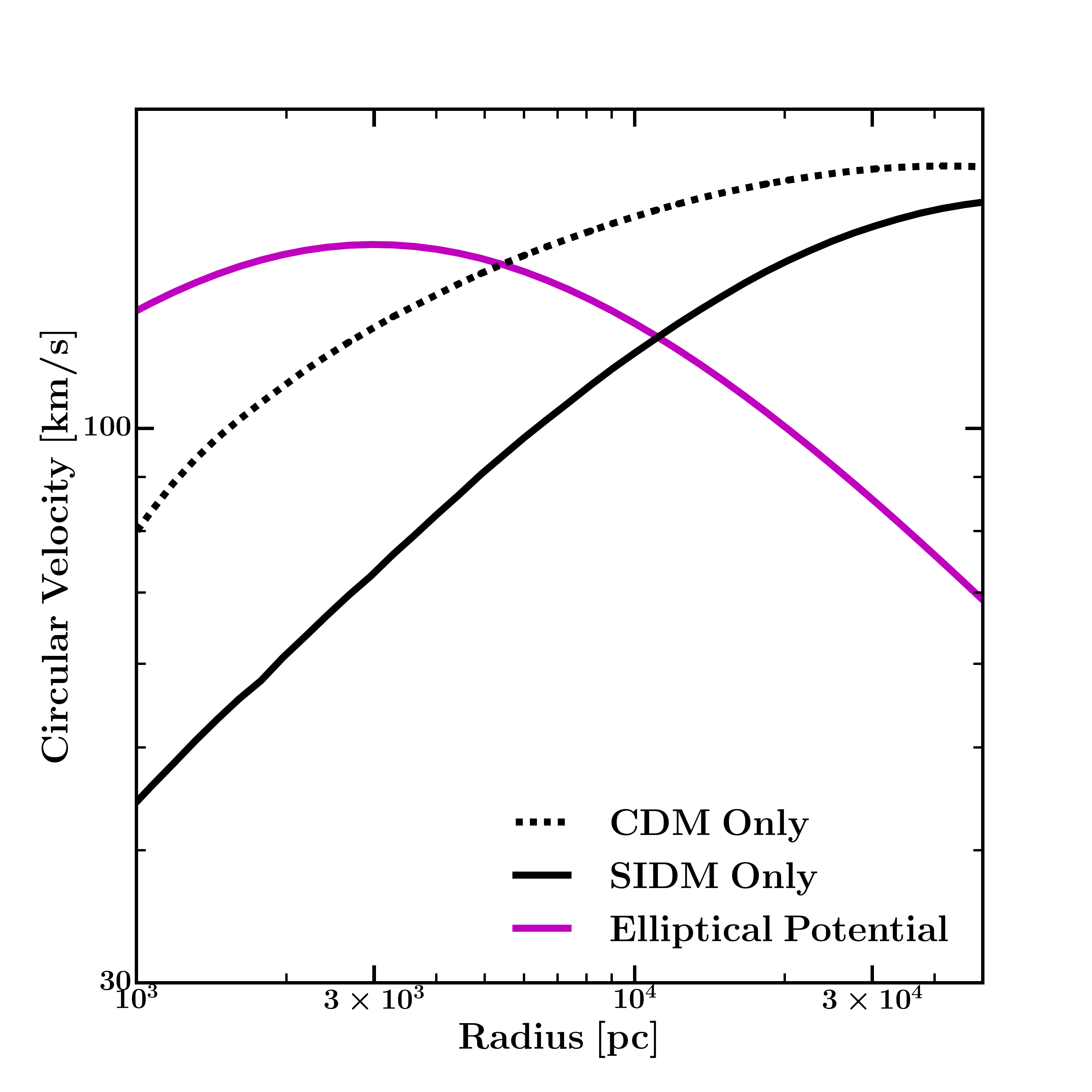} 
\includegraphics[width = \columnwidth]{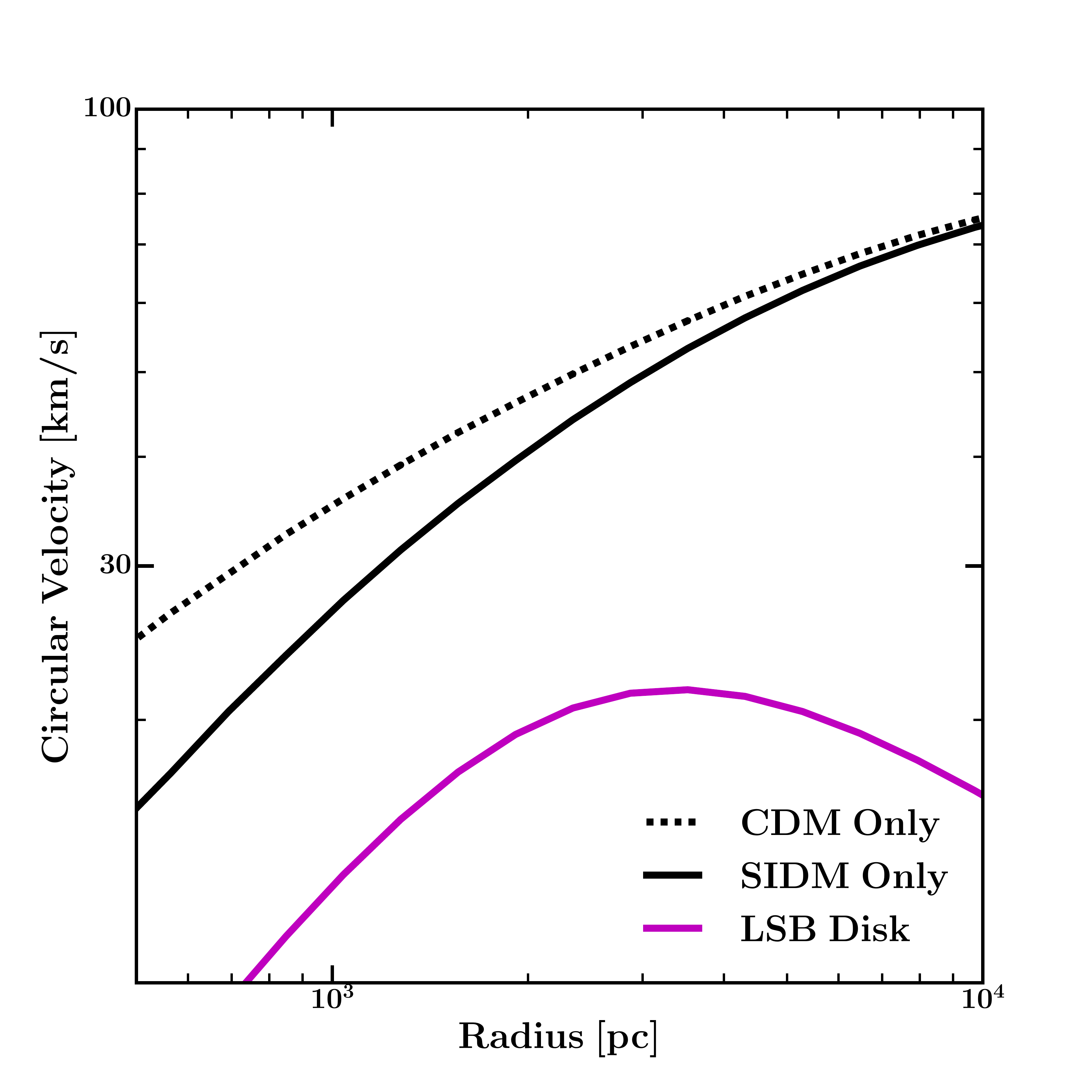} 
\hfill
\includegraphics[width = \columnwidth]{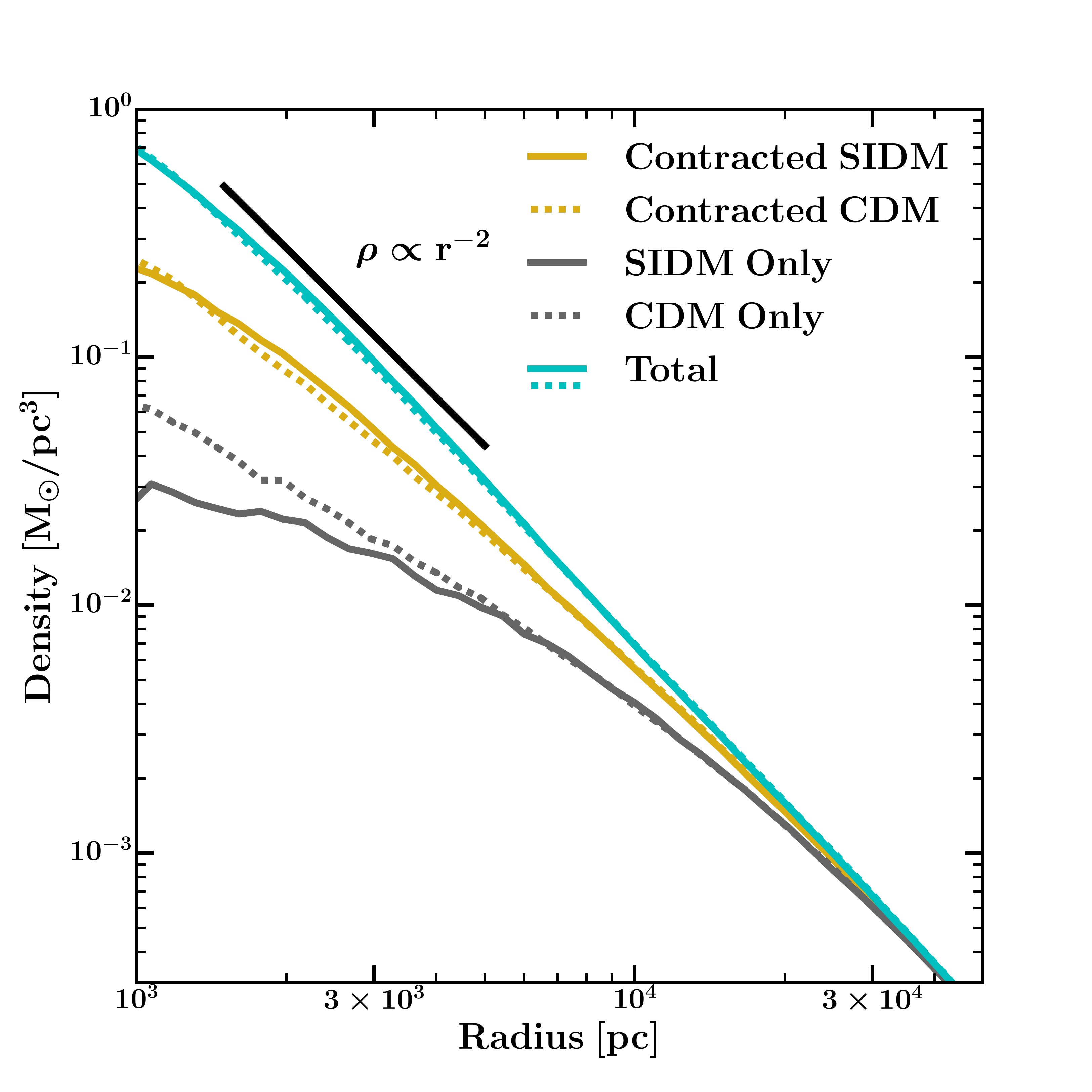} 
\includegraphics[width = \columnwidth]{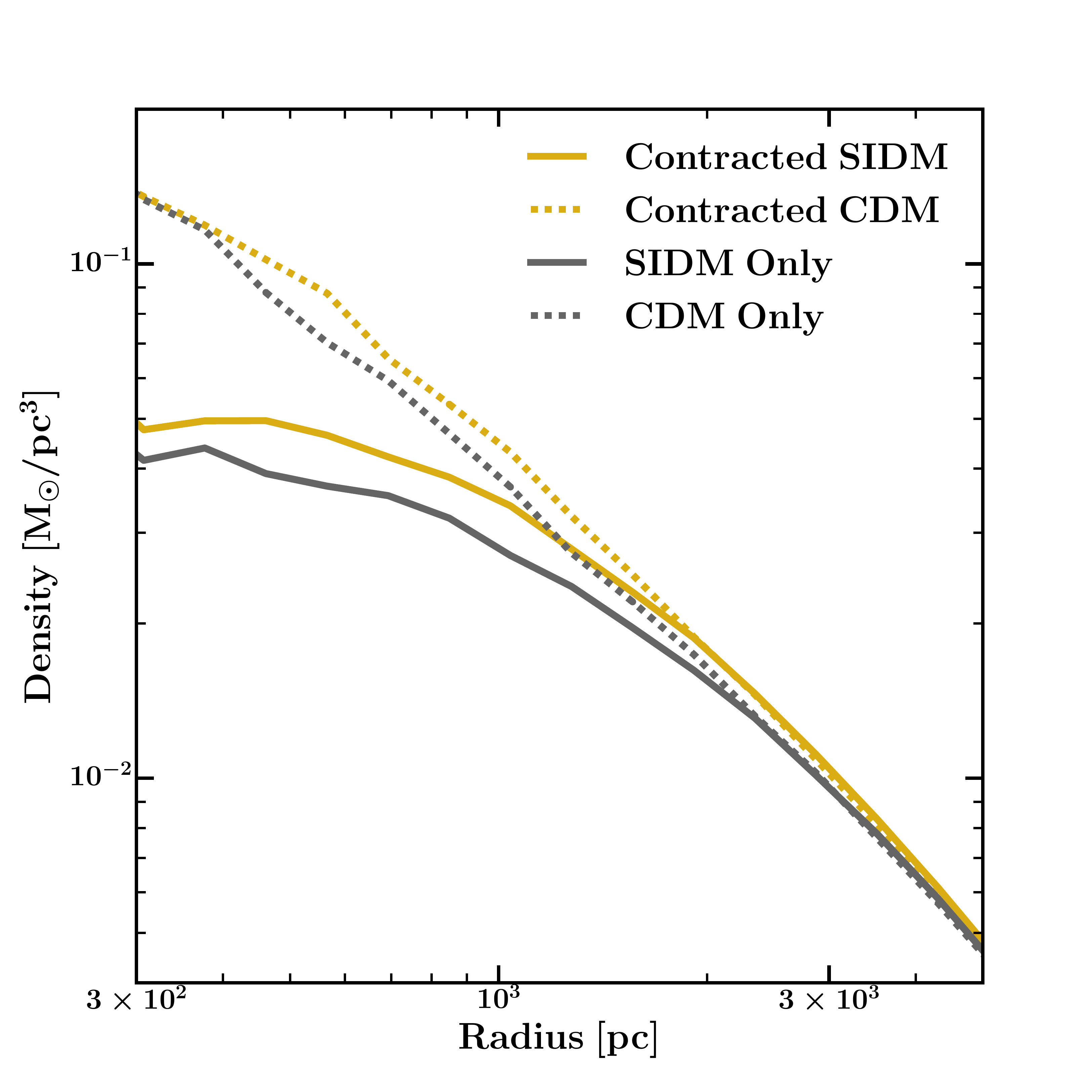} 
\caption{Initial rotation curves and final DM differential density profiles for our elliptical and LSB galaxy simulation simulated in both CDM and SIDM.  \textit{Left:} Elliptical Simulations.  The top panel shows the rotation curves for our CDM and SIDM halos as well as for the added Hernquist potential, displaying how the halo becomes baryon-dominated within 5 kpc.  The bottom panel shows the response of the halos to the added potential; as in the MW disk above, there is very little difference between the two contracted halos.  We also include the total (DM plus galaxy potential) density profiles in cyan, and an $\mathrm{r^{-2}}$ power-law for comparison.
\textit{Right:} LSB simulations.  The top panel shows the rotation curves for the initial CDM and SIDM halos along with the central disk potential, which contributes far less to the central mass and density of these halos.  The lower panel shows that this relatively shallow baryonic potential has a much smaller impact on the host halo; there is very little difference between the initial and contracted CDM halos.  The SIDM halo contracts slightly more, but still retains a core of lower density than the initial CDM halo.}
\label{fig:ellip}
\end{figure*}

Figure~\ref{fig:ellip} shows the density profiles and initial rotation curves of our elliptical and LSB halos. In the upper-left panel, we see that the elliptical halo is baryon-dominated within $\sim 5$ kpc, and consequently the density profiles (lower-left panel) show significant contraction.  Again we see minimal difference between the final SIDM and CDM halos, with no SIDM core resolved.  This is not surprising given our results from ~\ref{ssec:MW}; denser galaxies will have greater impacts on their host halos and inhibit SIDM core formation.  We also display the total (DM + baryons) density profiles in the lower-left panel of Figure~\ref{fig:ellip}, as well as an $r^{-2}$ power-law for comparison.  As a result of contraction, both simulations have power-law slopes of $\alpha=-2$ around 3 kpc, and slightly steeper outside this region.  This places our simulations in agreement with observations of elliptical galaxies \citep{Gavazzi07,Auger10}.  In short, because they are centrally baryon dominated, predictions for SIDM halos are largely the same as the CDM case for elliptical galaxies.

Our LSB-analogue simulations, however, do exhibit differences between CDM and SIDM runs.  
The right panel of Figure~\ref{fig:ellip} demonstrates that the LSB disk has very little effect on our dark matter halos: the CDM halo undergoes barely any contraction, and the SIDM halo still displays a central core.  This makes sense in light of our previous results that massive and centrally concentrated baryon densities generate the largest impacts on both SIDM and CDM halos. The diffuse nature of low-surface-brightness galaxies implies that their baryons have little effect on the host dark matter halos.  Thus, these systems are the best laboratories to investigate dark matter self-interaction possibilities.

\subsection{Analytic Model}
\label{ssec:analytic}

\begin{figure*}
\centering
\subfloat{\includegraphics[width = 0.7\columnwidth]{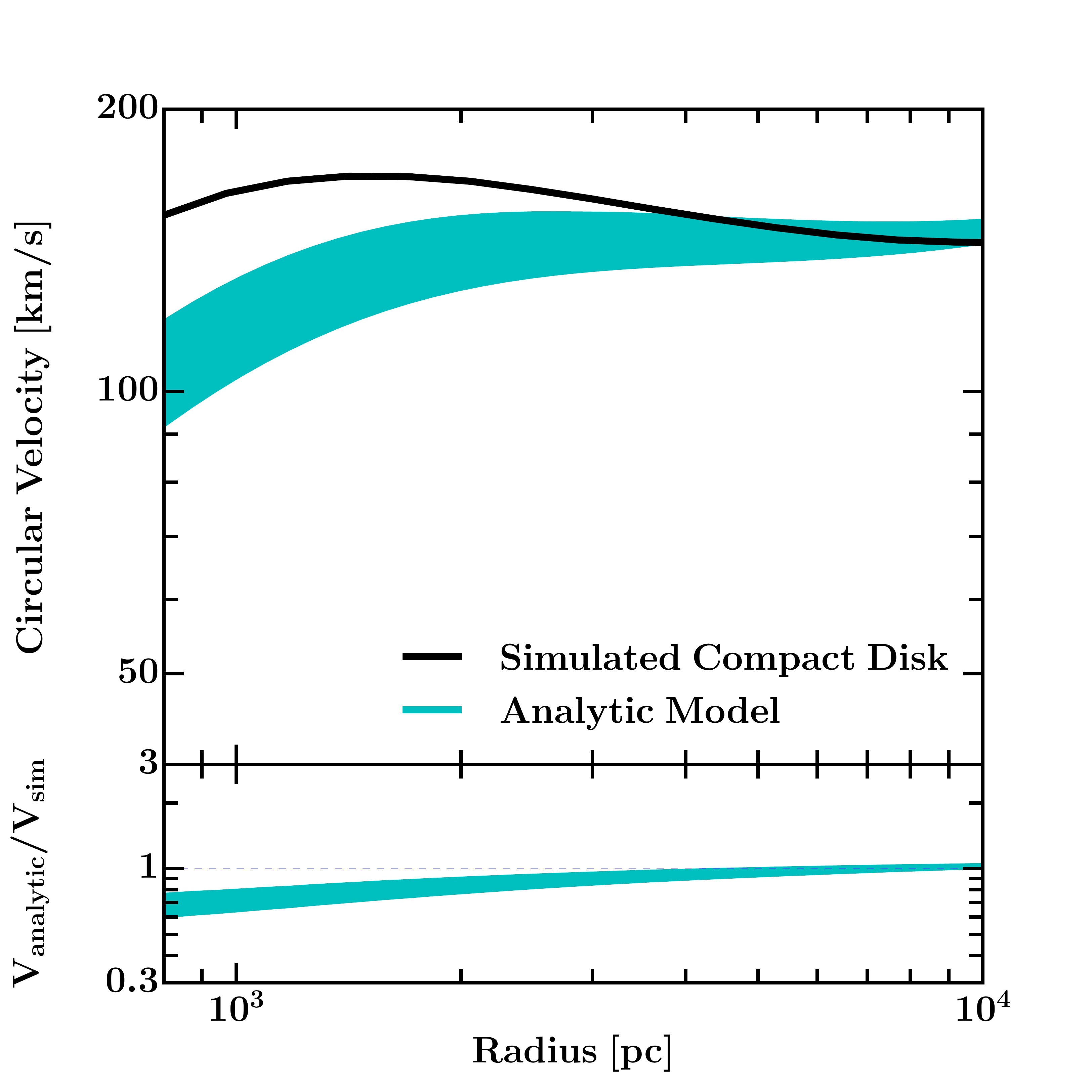}}
\subfloat{\includegraphics[width = 0.7\columnwidth]{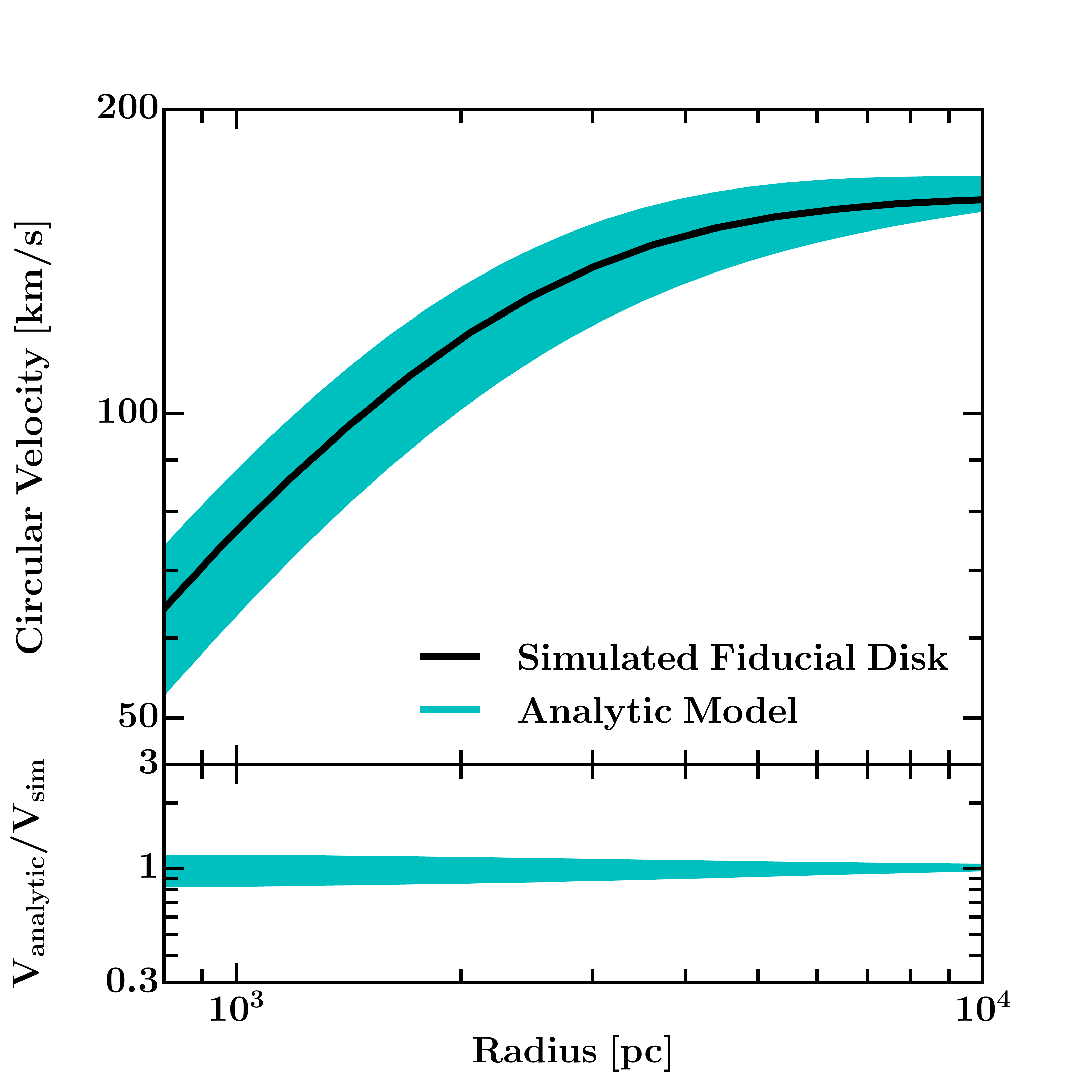}}
\subfloat{\includegraphics[width = 0.7\columnwidth]{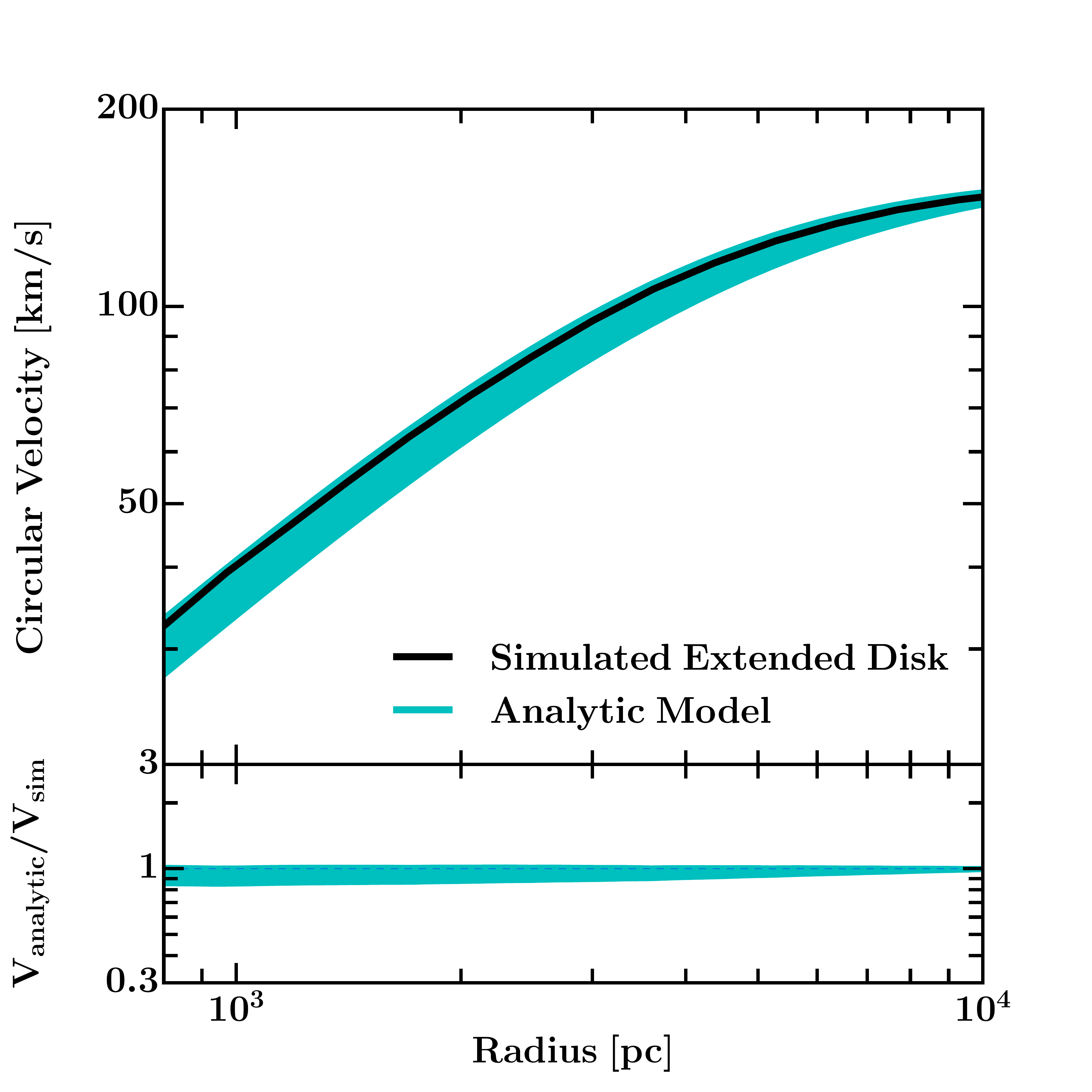}}
\hfill
\subfloat{\includegraphics[width = 0.7\columnwidth]{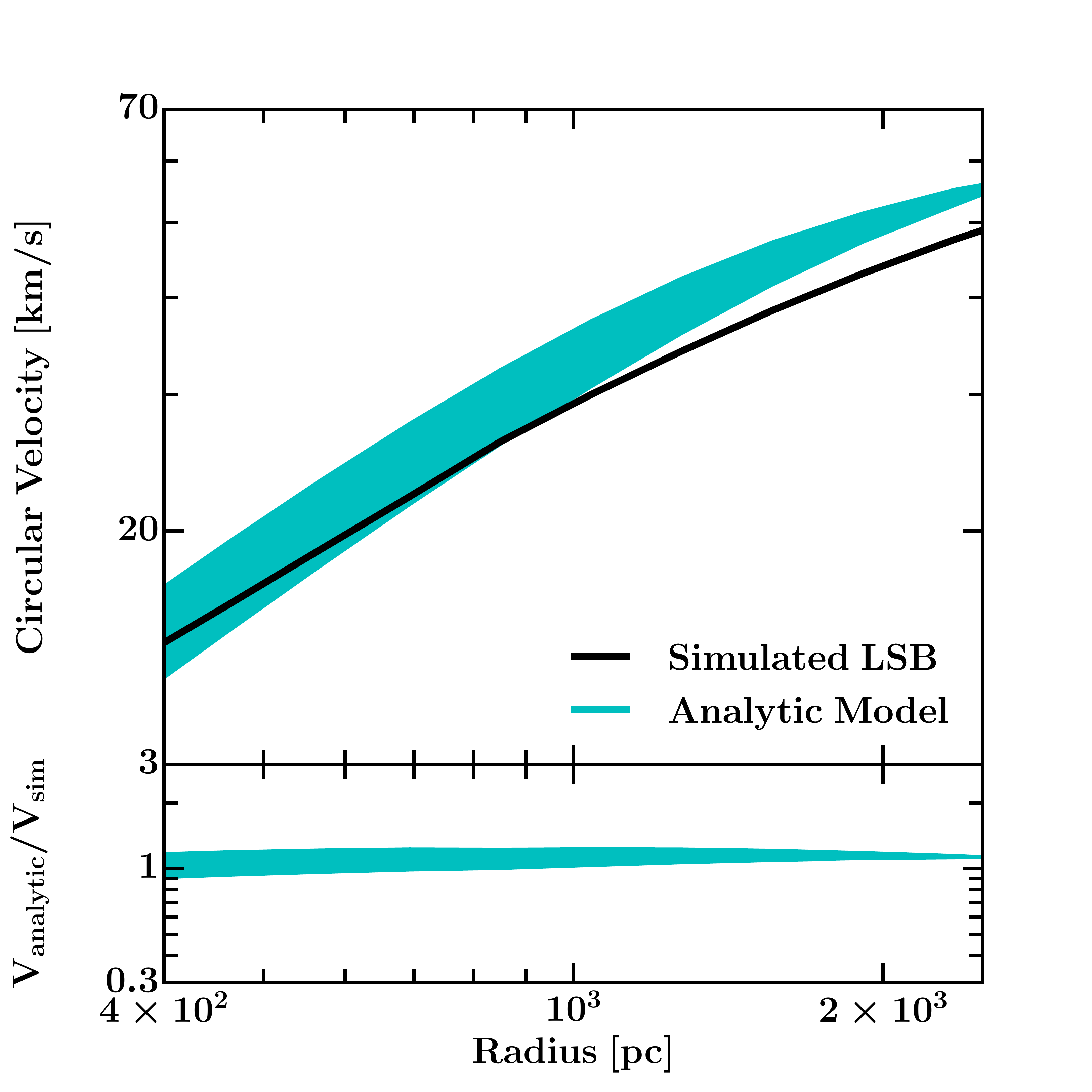}}
\subfloat{\includegraphics[width = 0.7\columnwidth]{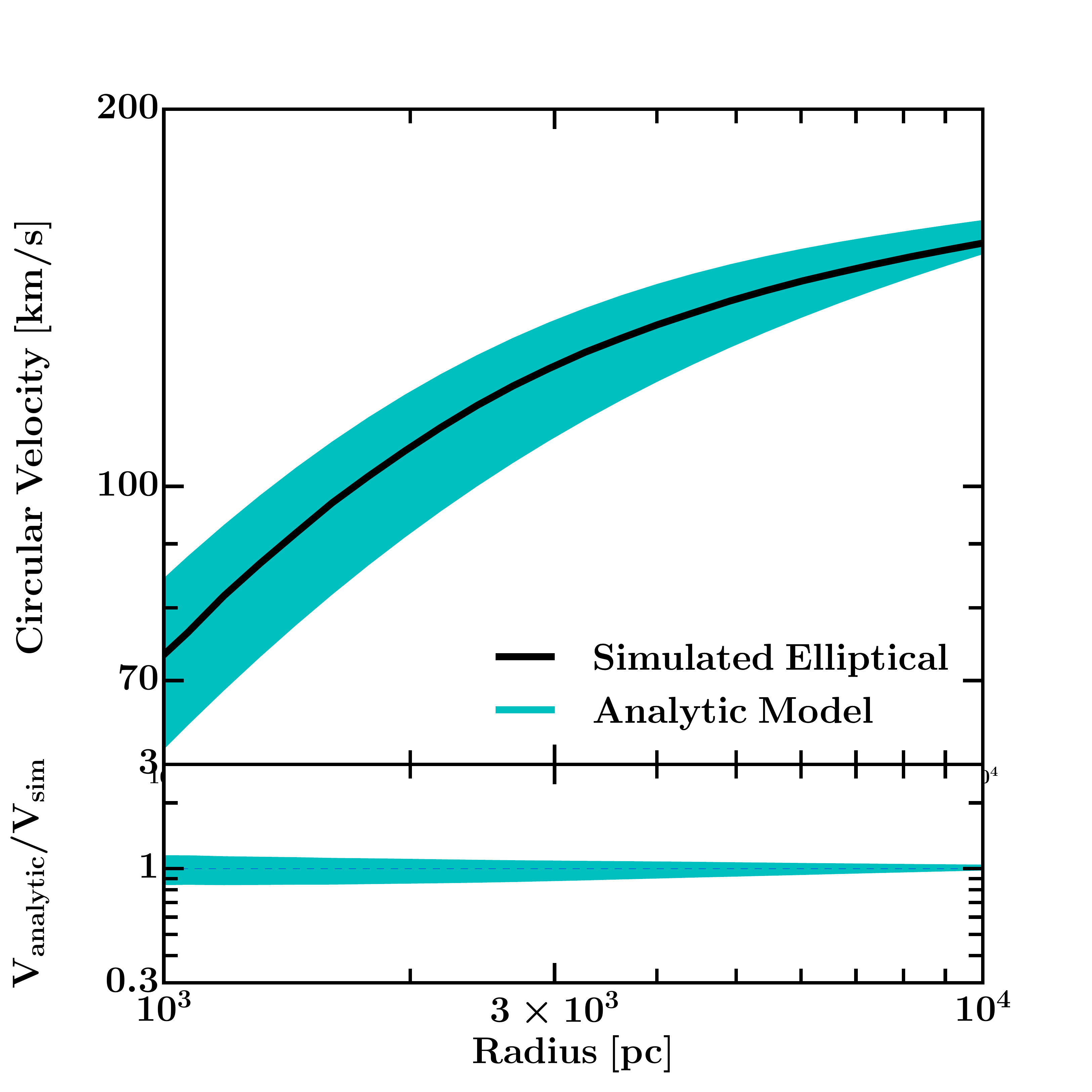}}
\subfloat{\includegraphics[width = 0.7\columnwidth]{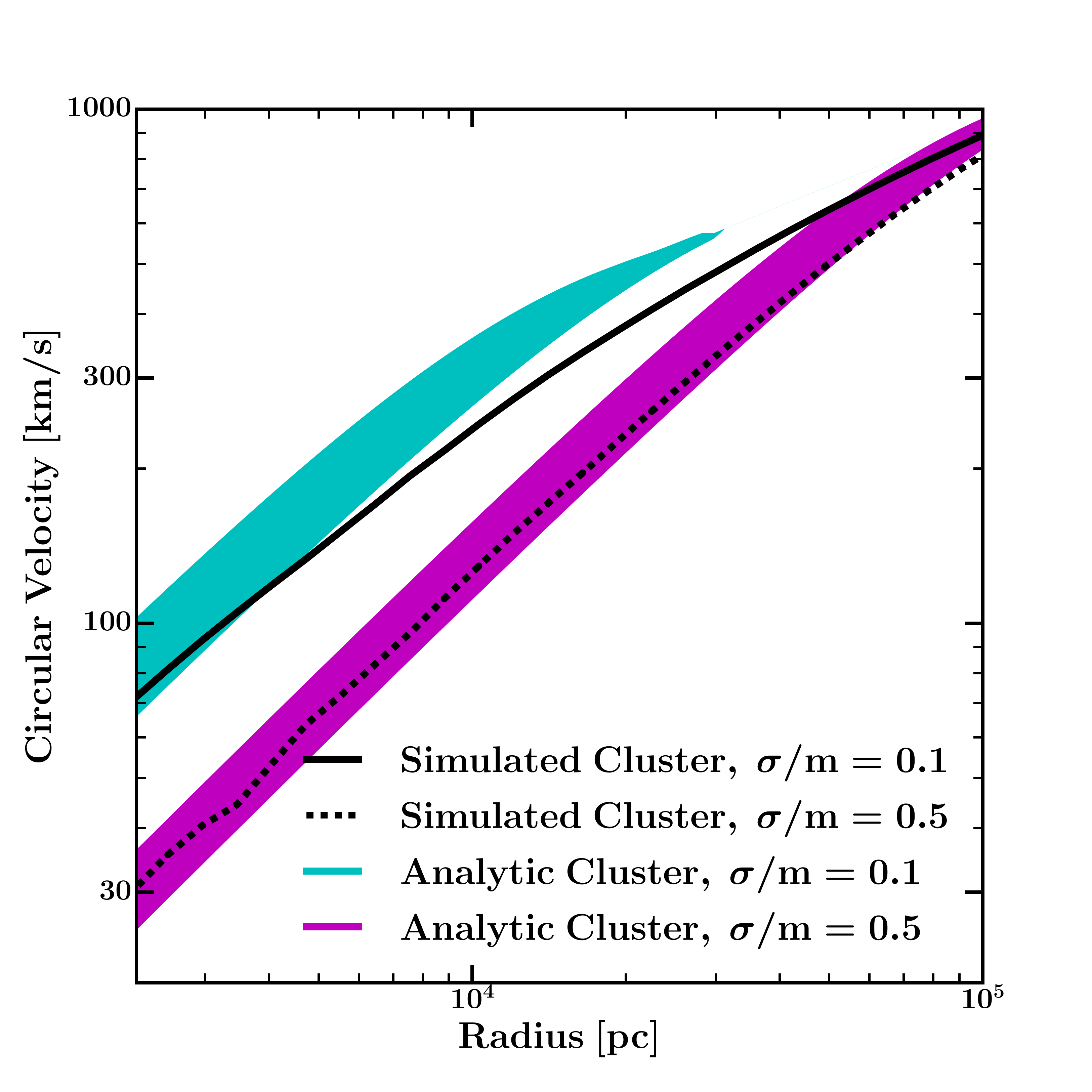}}
\caption{Comparison between circular velocities of our analytic model and simulations.  Generally the analytic model is in agreement with simulations, with the notable exception of the Compact Disk.  This is because of the core-collapse occurring in the Compact Disk simulation, causing greater central densities than predicted by our model, leading to higher circular velocities in the halo center.}
\label{fig:analytic}
\end{figure*}

In this section, we compare the results of our simulations to the analytic model presented in \citet{Kaplinghat15b}.  
In this model, the dark matter is assumed to behave as an isothermal gas within a radius $r_1$, defined as the radius where particles interact at least once in the age of the system: $\Gamma(r_1) \, t_\mathrm{age} = 1$ or
\begin{equation}
 \rho(r_1)\ (\sigmam)\ (4/\sqrt{\pi}) \sigma_{v0} \, t_\mathrm{age} = 1.
 \end{equation}
Here $\sigmam$ is the SIDM cross-section per unit mass, $\sigma_{v0}$ is the radial velocity dispersion in the core and  the factor of $4/\sqrt{\pi}$ accounts for the average relative velocity for Maxwellian distribution. $t_\mathrm{age}$ is the time period over which the self-interactions have been operating. We take this to be the time after the stellar component is fully in place to be consistent with the boundary conditions imposed. 

The analytic model we use is spherically symmetric. To apply it to our simulated galaxies, we follow the procedure of \citet{Kaplinghat14b} and construct a new {\em spherical} mass profile for the baryons $M_\mathrm{baryon}(r)$ by including all the mass in all the stars within a sphere of radius $r$.  Inside $r_1$ the SIDM density is set by hydrostatic equilibrium, giving 
\begin{equation}
\nabla^2 \ln \rho_\mathrm{DM} (r) =- \frac{4\pi}{\sigma_{v0}^2} \mathrm{G} \left[\rho_\mathrm{DM}(r)+\rho_\mathrm{baryon}(r)\right], 
\end{equation}
where $\rho_\mathrm{baryon}$ is the density profile corresponding to the mass profile $M_\mathrm{baryon}$.  At $r_1$, as boundary conditions, the isothermal mass and density profiles (from solving the hydrostatic equation above) are required to match the CDM halo profile after adiabatic contraction (essentially the $z=0$ CDM profile in our simulations).

Figure~\ref{fig:analytic} shows circular velocity profiles for our contracted SIDM halos along with the analytic predictions for rotation curves.  Here the cyan bands indicate the range of solutions which match $\rho(r_1)$ and $M(r_1)$ within 5\%, except for the core collapse case as discussed below.  We chose 5\% for two reasons. One, it shows the sensitivity of the inner density profile of SIDM profiles to the matching (boundary) conditions. Two, the simulated SIDM only and CDM only halos are not identical and examining a range of matching profiles makes us less sensitive to the differences in the simulated halo for the SIDM only and CDM only runs. 

Generally, the analytic model agrees well with the simulation results.  The glaring exception to this is the Compact Disk "Milky Way" run, for which the agreement is not good and 5\% matches were not found; for this run Fig.~\ref{fig:analytic} displays fits that match at the 15\% level.  This is because, as noted in Section~\ref{ssec:MW}, the halo hosting the compact disk is undergoing core-collapse, and so the assumption that the halo is isothermal is  violated, and the analytic solution returns a central density far smaller than observed in the simulations. Note that the fit is very sensitive to the value of the central dispersion $\sigma_{v0}$ because the inner density profile scales as $\exp(-\Phi_\star(r)/\sigma_{v0}^2)$. So, it should be possible to chose a lower central dispersion $\sigma_{v0}$ resulting in a higher inner dark matter density, but this would not be a useful exercise. In addition, we also found that our initial SIDM halos are about 0.1\% less massive than their CDM analogues at large radii, which could also throw off the analytic fit.  This difference in mass is constant beyond the SIDM core radius, implying that the mass removed from the halo centers in core formation escapes the inner halo, and potentially the halo itself.  This evaporation is not seen in cosmological simulations \citep[e.g.]{Rocha13} or captured in analytic treatment  \citep{Kaplinghat14b}, and is likely due to the truncated nature of our initial halo profiles.  Indeed, the difference is greatest in our cluster simulation, the most severely truncated halo.

The other main differences between the \citet{Kaplinghat15b} model predictions and our simulation results are in the profiles of the Cluster and LSB halos.  The circular velocities match well at small radii, but the simulated profiles are flatter than the analytic ones.  We believe this is primarily caused by the halo evaporation mentioned above. If so, this is a numerical artifact of the halo truncation adopted in our initial conditions and not a prediction of the SIDM models.  In our other simulations where the stellar gravitational potential is dominant enough, this mass loss is not seen.

Aside from these differences, the analytic model does an excellent job predicting the densities and central circular velocities of the contracted halos. The success of the model for the Milky Way analogs is surprising because the assumption of spherical symmetry in the analytic model is grossly violated. Given the overall consistency of the analytic model predictions with our simulation halo profiles, we are confident that the \citet{Kaplinghat15b} model treats contraction in SIDM halos due to baryons well. In particular, it may be used to model SIDM halos with cross sections other than that simulated here (0.5 $\cmg$).  

\section{Cluster Limits}
\label{sec:cluster}

\begin{figure*}
\centering
\subfloat{\includegraphics[width = .715\columnwidth,trim={1.2cm 0 1cm 0}]{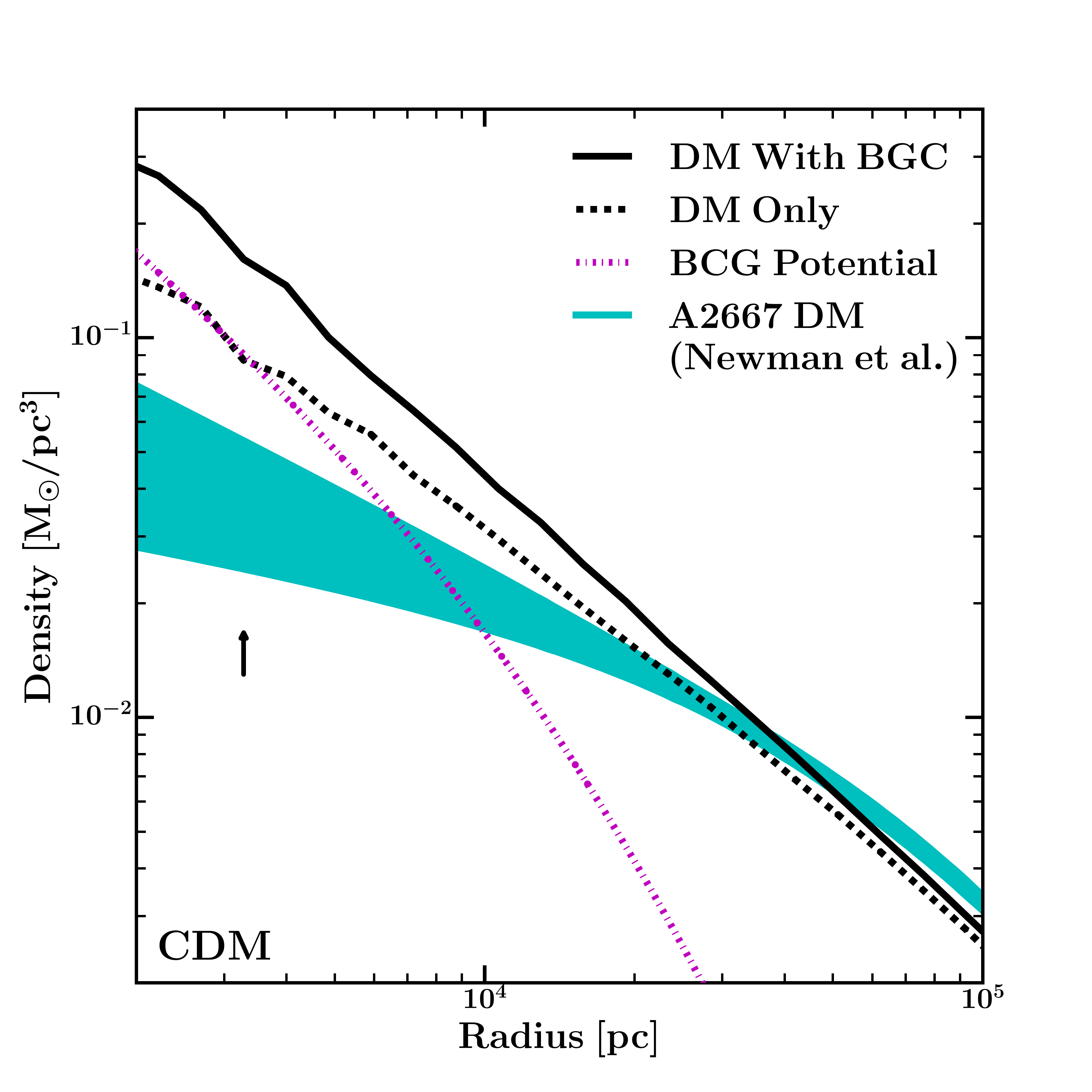}}
\subfloat{\includegraphics[width = .715\columnwidth,trim={1.2cm 0 1cm 0}]{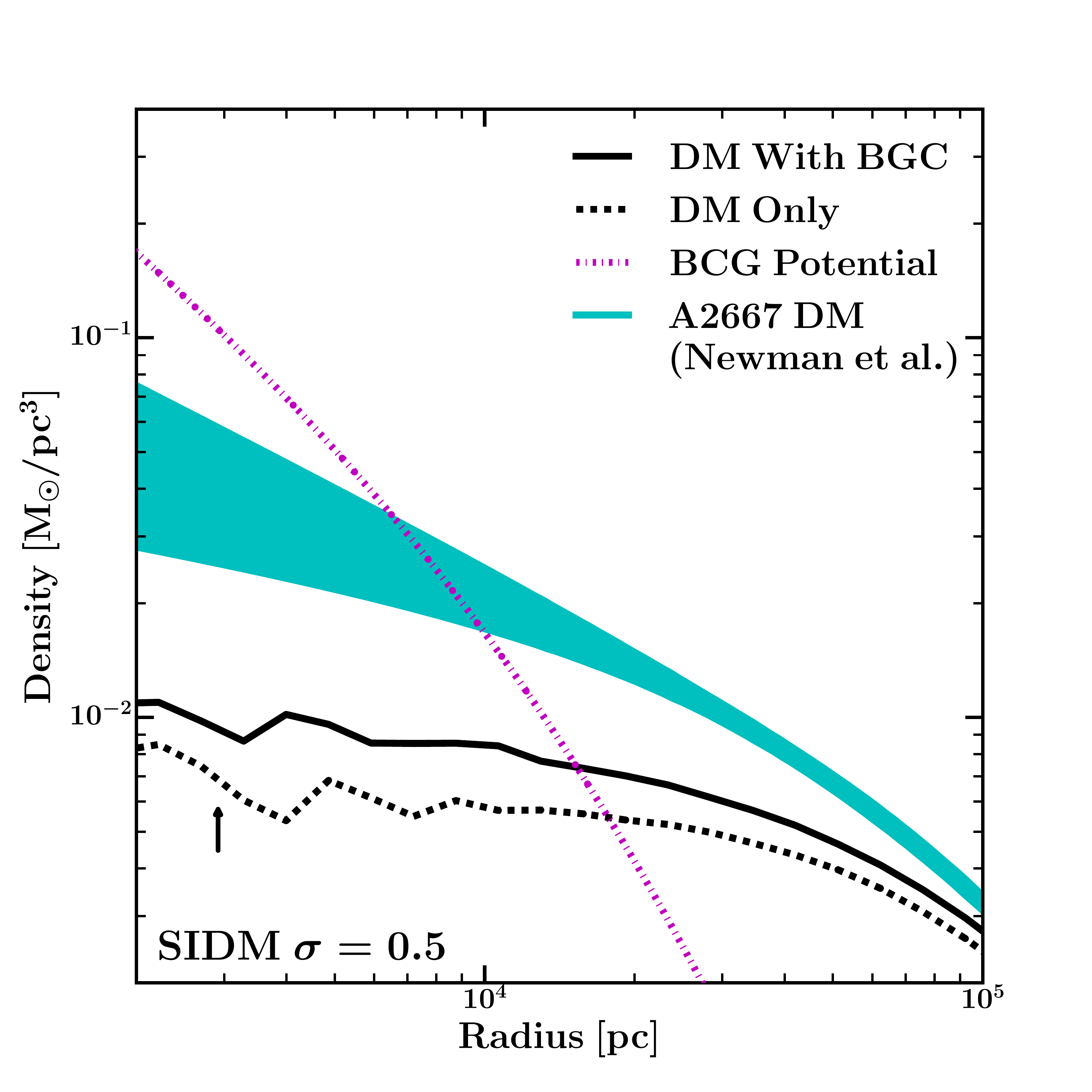}}
\subfloat{\includegraphics[width = .715\columnwidth,trim={1.2cm 0 1cm 0}]{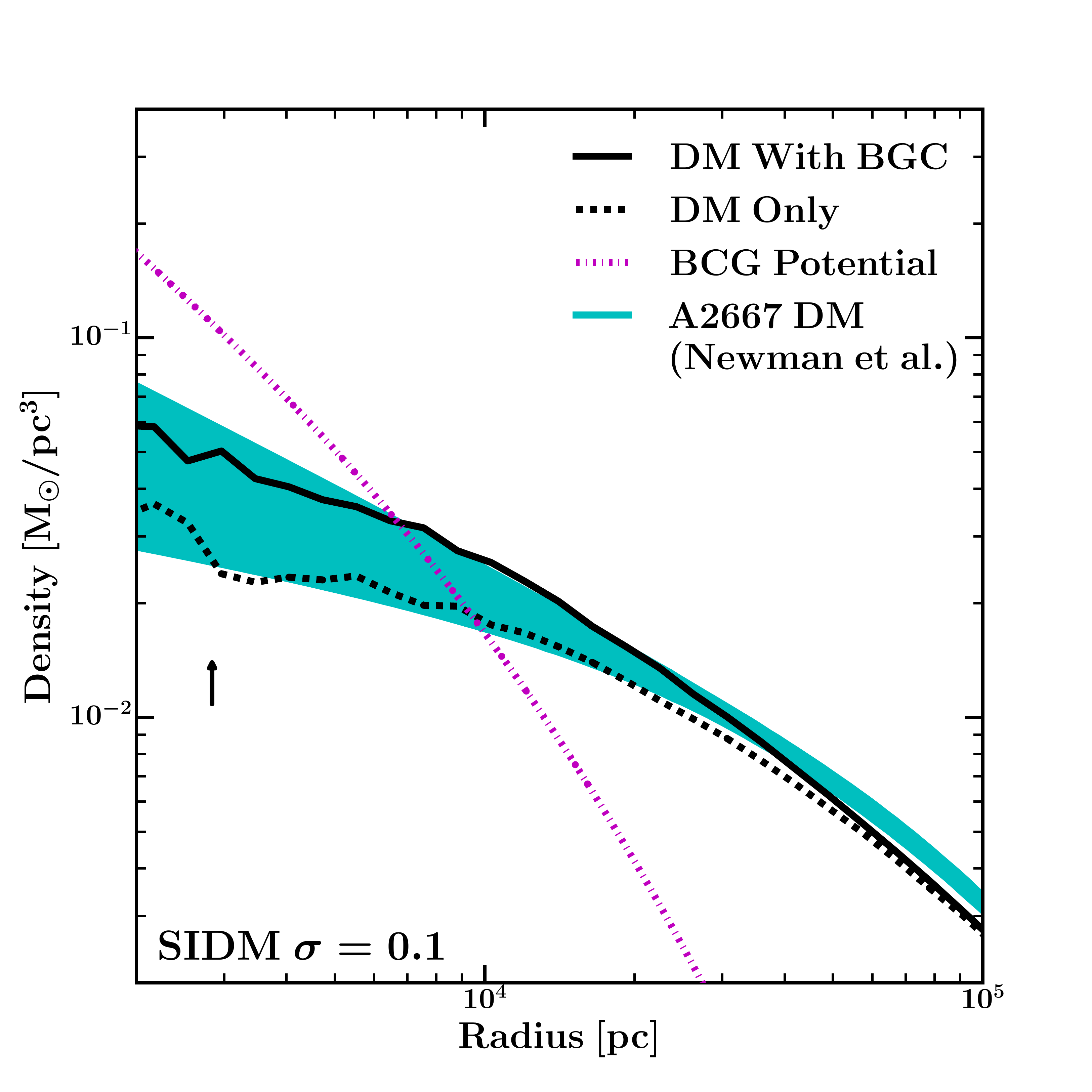}}
\caption{Density profiles for our simulations of cluster Abell 2667, along with the inferred dark matter density profile from \citet{Newman13b}. Even before contraction, our CDM halo (left) is too dense and the contracted SIDM profile with $\sigmam=0.5\ \cmg$ (center) is not dense enough to reproduce the observed density profile.  Our contracted SIDM halo with $\sigmam=0.1\ \cmg$ (right), however, matches the observed density of Abell 2667 well, in agreement with the predictions of \citet{Kaplinghat15b}.}
\label{fig:cluster}
\end{figure*}

We also simulate an analog of Abell cluster 2667.  Our setup is described in the bottom row of Table~\ref{sims.tab}.  The virial mass and concentration of the initial NFW halo match the observations of \citet{Newman13a}, however they model the density of the BCG as a dual psuedo-isothermal elliptical profile (dPIE): $$\rho=\frac{\rho_0}{1+r^2/(r_\mathrm{core}^2)(1+r^2/r_\mathrm{cut}^2)}$$  

In order to model this in our simulations we use a \citet{Hernquist90} sphere with the same half-mass radius as the measured profile, and use a least-squares fitting method to obtain a best-fit total mass.  This is necessary because dPIE profiles exhibit central cores, while Hernquist profiles have central $1/r$ cusps, so a Hernquist distribution with the same mass and half-mass radius as a dPIE distribution will be less dense at all radii outside of the core radius, significantly reducing the contraction effect we wish to investigate.  We grow the central potential over the course of 2 Gyr, and allow the halo to evolve for 3 Gyr after this, running for 5 Gyr total.

Figure~\ref{fig:cluster} shows the density profiles of our simulations; uncontracted densities are plotted as black dashed lines, while the contracted halos are solid black.  We also include a vertical arrow to indicate the smallest converged radius of our simulations, as defined in \citet{Elbert15}.  We plot the inferred Cored NFW dark matter density profile measured by \citet{Newman13b} as a cyan band, though we note that the generalized NFW fits \citet{Newman13b} reported are not significantly different.  Finally, the effective density of the BCG Hernquist profile is shown in magenta.  The leftmost figure displays our CDM simulations.  Even before contraction, the initial NFW profile is too centrally dense to match observations by a factor of $\sim 1.5$ at the limits of our resolution.  Adiabatic contraction increases the density by another factor of $1.5$, further worsening this discrepancy.  If instead we assume the dark matter is self interacting with a cross section of $0.5\ \cmg$ we obtain the densities in the center plot.  In this case, we see that such a cross-section results in a halo that is underdense by more than a factor of 3 in its center, and still below the observed limits at 100 kpc in our SIDM-only simulation.  While baryonic contraction does increase the DM density, it is not nearly strong enough to completely alleviate the problem.  Indeed, over most of our resolved region the contracted halo is only $20-30\%$ higher than the SIDM-only simulation.

This result may seem surprising at first given that the baryons are important in the center in terms of their dynamical mass. However, the key point is that the DM halo is so massive that its velocity dispersion dwarfs the stellar potential. In this respect, the contraction of the SIDM halo is different from the adiabatic contraction of CDM halos; applying the \citet{Blumenthal86} adiabatic contraction formula would result in a halo profile much denser than the simulated result. The isothermal equilibrium solution, on the other hand, is a good match to the simulated halo profile. 

We also simulate Abell 2667 with a cross-section of $0.1\ \cmg$; our results are plotted in the right-panel of Figure~\ref{fig:cluster}.  In the SIDM only case, the density is near the lower-limits of the \citet{Newman13b} data.  After contraction has been accounted for, however, $0.1\ \cmg$ is consistent with observations, bordering the upper limits of the \citet{Newman13b} measurement, and we estimate that a cross section of $0.2\ \cmg$ will border the lower limits. Our estimates are in agreement with the results of \citet{Kaplinghat15b}, who find cross sections of $0.1-0.2\ \cmg$ (assuming $t_\mathrm{age} = 5$ Gyr) by fitting to the \citet{Newman13b} data using the analytic model (described previously). This implies that a SIDM model with a cross section that falls from about $1\ \cmg$ on dwarf galaxy scales to about $0.1\ \cmg$ on cluster scales can resolve the small-scale puzzles  \citep{Kaplinghat15b,Elbert15,Vogelsberger14,Zavala13}, while also matching density profile constraints in clusters.

\section{Conclusions}
\label{sec:conclusions}

In this work we have investigated the combined effects of baryonic gravitational potentials and dark matter self-interactions on dark matter halos using idealized simulations of dark matter halos with galactic potentials.  By simulating halos of various sizes with many different potentials we have found:

\begin{itemize}
\item SIDM halo shapes are not inherently more resilient to effects from baryons than their CDM counterparts.  For a Milky Way halo hosting a Milky Way analogue disk, the SIDM halo is more compact along the disk axis than its CDM equivalent in agreement with the prediction of \citet{Kaplinghat14b}. For an elliptical galaxy, whose stellar potential is markedly more spherical, we expect the SIDM halo to be correspondingly more spherical.  

\item Halos that host substantial baryonic populations display few differences in spherically-averaged density profiles between CDM and SIDM models on observable scales.  Even extended baryon populations can contract halos with respect to SIDM-only simulations, though these systems retain potentially observable constant-density cores and are less dense than CDM.  In extreme cases, we find that potentials from dense baryonic structure can cause SIDM halos to core-collapse and become denser than their CDM counterparts. 

\item Halos that host less massive or highly diffuse stellar and gas disks display substantial differences between CDM and SIDM models.  Thus, the original motivation for explaining the low densities observed in galaxies dominated by dark matter is still intact. Among galaxies, these are likely the best systems to measure or constrain the SIDM cross section.  

\item The densities of our contracted halos are in good agreement with the analytical predictions in \citet{Kaplinghat14b, Kaplinghat15b}, with the exception of the core-collapsing Compact Disk because it no longer obeys the isothermal assumption of the model. In particular, we show that the spherically-averaged density profiles are well approximated by the simple model in \citet{Kaplinghat15b}, which has an isothermal core and an undisturbed CDM outer profile. 

\item We simulated a cluster halo for 3 Gyr after the brightest cluster galaxy was fully in place to test against the mass measurements for A2667 \cite{Newman13b}. Our simulated CDM halo was denser than the observed central profile for A2667.  On the other hand, SIDM with $\sigma/m \simeq 0.5\ \cmg$ was too low in density compared to the measurements.  The choice of $\sigma/m \simeq 0.1\ \cmg$ was in much better agreement with the measured normalization (and inner slope) of the A2667 density profile. 
Larger values like $\sigma/m \simeq 0.5\ \cmg$ are ruled out, even allowing for a factor of 2 uncertainty in the age of the halo. These conclusions are in substantial agreement with the detailed analysis of seven clusters \citep{Newman13b} by \citet{Kaplinghat15b}, which found a average value of $\sigma/m \simeq 0.1\ \cmg$ on cluster velocity scales for an assumed age of 5 Gyr. 
\end{itemize}

Based on these results, an ideal scale to investigate possible DM self-interactions appears to be the dwarf galaxy scale with halo masses $10^{10-11}\ \msun$ scale, as they will have the largest interaction cross sections and the least contracted halos.  However, these are precisely the halos expected to be most vulnerable to stellar feedback \citep{Pontzen12,Governato12,DiCintio14,Onorbe15}.  Ongoing work \citep{Vogelsberger14,Fry15,Roblesinprep} is investigating the effects of feedback on the SIDM halos and their results suggest that dwarfs with stellar masses $M_\star \lesssim 10^6 \msun$ will have density profiles indistinguishable from the predictions of the dark matter-only simulations. This suggests that the faintest dwarf spheroidals provide excellent laboratories constraining SIDM models.   

For halo masses much larger than $10^{11}\ \msun$ that host a large stellar disk or bulge, as the inner halo becomes isothermal the SIDM halo retains the high densities created by adiabatic contraction following the formation of the disk.  Thus, in Milky Way sized halos the CDM and SIDM halos densities are very similar beyond about a kpc, in marked contrast to the dark-matter-only predictions. As predicted by \citet{Kaplinghat14b}, the self-interactions also force the SIDM halo to be more compact along the stellar disk axis. We find that the SIDM halo in the inner region of Milky Way analogs is more compact along the galactic disk axis than the CDM halo. Thus, it may be possible to use the shape of the dark matter halo in the inner regions of large spiral galaxies to provide a sharp test of the SIDM paradigm. 

The predictive cross-talk between the dark matter and baryons in the SIDM models leads to a large diversity of halo profiles and halo shapes. This cross-talk is purely gravitational and the result of the dark matter becoming isothermal in the inner parts of the halos and they are fully explained by simple equilibrium models. The prospects for using these concrete predictions of the SIDM paradigm to rule in or rule out SIDM in the near future are excellent. 

\section*{Acknowledgements}

The authors thank Michael Boylan-Kolchin and Hai-Bo Yu for valuable discussions.
ODE, ASG, and JSB were supported by the National
Science Foundation (grants PHY-1520921, AST-1518291, and AST-1009973) and by
NASA through HST (GO-13343) awarded by the Space Telescope Science Institute
(STScI), which is operated by the Association of Universities for Research in Astronomy
(AURA), Inc., under NASA contract NAS5-26555.
MK is supported by NSF grant PHY-1620638.
Support for SGK was provided by NASA through Einstein Postdoctoral Fellowship
grant number PF5-160136 awarded by the Chandra X-ray Center, which is operated
by the Smithsonian Astrophysical Observatory for NASA under contract NAS8-03060.

This work also made use of \texttt{Astropy}, a community-developed core Python package
for Astronomy \citep{Astropy}, \texttt{matplotlib} \citep{Matplotlib}, \texttt{numpy}
\citep{numpy}, \texttt{scipy} \citep{scipy}, \texttt{ipython} \citep{ipython}, and NASA's
Astrophysics Data System.

\nocite{xsede}
\bibliography{SIDM}

\label{lastpage}
\end{document}